\documentclass[12pt,letterpaper]{article}
\usepackage[T1]{fontenc}
\usepackage[utf8]{inputenc}
\usepackage{lmodern,textcomp,microtype}
\IfFileExists{parskip.sty}{\usepackage{parskip}}{\setlength{\parindent}{0pt}\setlength{\parskip}{6pt plus 2pt minus 1pt}}
\usepackage{amsmath,amssymb,amsthm,mathtools}

\usepackage{booktabs,longtable,array,graphicx,placeins}
\usepackage[font=small,labelfont=bf]{caption}
\usepackage[authoryear,round]{natbib}
\usepackage[hidelinks]{hyperref}
\IfFileExists{xurl.sty}{\usepackage{xurl}}{\usepackage{url}}
\newtheorem{theorem}{Theorem}
\newtheorem{proposition}{Proposition}

\theoremstyle{remark}

\newcommand{\R}{\mathbb{R}}

\def\spacingset#1{\renewcommand{\baselinestretch}{#1}\small\normalsize}
\hypersetup{pdftitle={The Lambert Penalty: Logarithmic Shrinkage for Sparse Regression},pdfauthor={Bahadır Yüzbaşı},pdfcreator={LaTeX}}

\begin{document}
\spacingset{1}
\title{\bfseries The Lambert Penalty: Logarithmic Shrinkage for Sparse Regression}
\author{Bahad{\i}r Y\"uzba\c{s}{\i}\\[3pt]
Department of Econometrics, \.{I}n\"on\"u University,\\
44280, Malatya, T\"urkiye\\[3pt]
\small Correspondence: \href{mailto:bahadir.yuzbasi@inonu.edu.tr}{bahadir.yuzbasi@inonu.edu.tr}\\
\small ORCID: \href{https://orcid.org/0000-0002-6196-3201}{0000-0002-6196-3201}}
\date{}
\maketitle
\begin{abstract}
Sparse regression must balance prediction accuracy with reproducible variable
selection. We introduce the Lambert penalty by limiting how quickly the
retained fraction of a scalar score increases after variable entry.
Maximizing retention under this scale-equivariant constraint yields a
logarithmic transition from exact zero to an identity tail. The bounded
penalty has a fixed shape and a single penalty parameter selected by
training-only five-fold cross-validation. We derive an explicit scalar
update, a sharp weak-convexity bound, a sufficient condition for multivariate
uniqueness, and conditional convergence of exact cyclic coordinate descent.
For fixed dimension, global minimizers are uniformly equivalent to the oracle
estimator over an asymptotic tuning band. Gaussian simulations compare
Lambert with LASSO, ridge, elastic net, MCP, and SCAD. Lambert favors
prediction for strong equal signals and usually improves support recovery
over SCAD; MCP often gives better support recovery, and convex methods remain
competitive with weak signals or strong correlation. In a QSAR toxicity
application, repeated nested cross-validation gives competitive prediction
with compact, stable models, although error differences from MCP and SCAD
are small. The empirical design followed exploratory development; independent
confirmation remains necessary.
\end{abstract}
\newpage
\spacingset{1}
\section{Introduction}
\label{sec:introduction}

Sparse regression requires decisions about both which variables to retain
and how strongly to shrink their estimated effects. These decisions are
linked: shrinkage can stabilize estimation, but persistent shrinkage of
substantial effects introduces bias, whereas rapidly releasing shrinkage
may increase sensitivity to noise. A useful penalty must therefore specify
more than an entry threshold. It must also determine how estimated effects
grow between exclusion and negligible shrinkage. That transition matters
for prediction and support recovery, whose preferred solutions need not
coincide \citep{fanli2001,zhang2010}.

Established penalties resolve this tradeoff in different ways. The LASSO
produces exact zeros through a convex objective, but its constant penalty
slope continues to shrink large coefficients \citep{tibshirani1996}. Elastic
net combines absolute-value and quadratic penalties to stabilize estimation
and selection under correlated predictors \citep{zouhastie2005}. SCAD and MCP
reduce the penalty slope as coefficient magnitude increases and eventually
leave sufficiently large scalar scores unchanged \citep{fanli2001,zhang2010}.
These constructions show how sparsity can coexist with reduced bias for
large effects. Their intermediate transitions, however, impose different
patterns of shrinkage on effects that have entered the model but are not
yet fully retained.

A broader literature constructs penalties from specified shrinkage mappings
\citep{chartrand2014,woodworthchartrand2016}. Smooth SCAD-type rules also
provide nonlinear transitions between exact zeros and an identity tail
\citep{kulkarni2026smoothscad}. Thus neither construction from a shrinkage
rule nor a nonlinear transition is new. The design question pursued here
is more specific: how can the release of shrinkage be controlled on a
relative scale, while retaining continuous entry, a finite identity cutoff,
and a fixed shape with only one tuning parameter?

We address this question by limiting the increase in the retained fraction
of a scalar score per unit increase in its logarithm. A logarithmic scale
treats proportional changes in score magnitude alike: a common rescaling
of the score and penalty level leaves the restriction unchanged. Within
this class, we choose the pointwise largest feasible retention rule. The
unit rate used in the main method makes its entry slope equal to that of
soft thresholding; thereafter the retained effect grows increasingly
quickly until it reaches the unpenalized score. This construction gives a
logarithmic transition whose inverse, expressed through the principal
Lambert function, defines the bounded Lambert penalty. The rate is fixed
by design, and only the penalty strength is selected by cross-validation.
Section~\ref{sec:retentiondesign} states the constraint and its solution.
Maximal retention within this class is a design property; it does not by
itself imply minimum statistical risk.

The contribution combines this specified retention constraint with an
explicit penalty and an analysis of the resulting regression estimator.
Section~\ref{sec:audit} distinguishes its derivatives and scalar update
from related constructions. We obtain the exact scalar minimizer and a
sharp weak-convexity bound, which identify the coordinate update and quantify
the penalty's contribution to objective curvature. We then establish a
sufficient condition for multivariate uniqueness and conditional convergence
of exact cyclic coordinate descent. For fixed dimension, we prove uniform
equivalence between global minimizers and the oracle least-squares estimator
over an asymptotic tuning band. This last result concerns global minimizers
under its tuning assumptions; it does not assert oracle behavior for every
cross-validation choice or stationary fit.

The empirical study examines when this shrinkage pattern is useful.
Section~\ref{sec:fixedempirical} compares Lambert with LASSO, ridge, elastic
net, MCP, and SCAD across signal strengths, heterogeneity, and predictor
correlations. Prediction, coefficient estimation, and support recovery are
assessed separately. Strong signals of equal magnitude provide a favorable
prediction regime for Lambert, while support recovery and weak-signal
performance reveal tradeoffs with competing methods. Section~\ref{sec:qsar}
presents a QSAR aquatic-toxicity application, where the true support is
unknown and predictive accuracy is evaluated alongside model size and
selection stability. The accompanying \texttt{lambertReg} package provides
regularization paths, cross-validation, prediction, and diagnostic tools.

\section{The Lambert penalty and estimator}
\label{sec:model}

Let $\lambda>0$ be the penalty strength and let $t\geq0$ denote an absolute
coefficient. Write $W_0:[0,\infty)\to[0,\infty)$ for the principal real
Lambert function, characterized by $W_0(u)e^{W_0(u)}=u$. With $e$ denoting
the base of the natural logarithm, define the Lambert penalty explicitly by
\begin{equation}
 p_\lambda(t)=
 \begin{cases}
 \displaystyle\frac{\lambda^2}{4}
 \left[e^{2w}(1+2w-2w^2)-1\right],\quad w=W_0(t/\lambda),
       &0\leq t\leq e\lambda,\\[5pt]
 \displaystyle\frac{e^2-1}{4}\lambda^2,&t>e\lambda.
 \end{cases}
 \label{eq:mainpenalty}
\end{equation}
Equation~\eqref{eq:mainpenalty} is the penalty used throughout the main
analysis. It is zero at the origin, increases to a finite plateau, and has
zero slope beyond $e\lambda$. It is the $c=1$ member of the broader family
in Supplementary Section~\ref*{sec:family}; that shape parameter is not tuned. We use
$p_\lambda$ and the name Lambert without repeating the fixed shape below.

For centered responses $y\in\R^n$ and a centered design matrix
$X\in\R^{n\times p}$, standardize each column $X_j$ to satisfy
$X_j^{\mathsf T}X_j/n=1$. The vector $\beta\in\R^p$ denotes coefficients on
this standardized scale. Define the statistical estimator as any global
minimizer of
\begin{equation}
 Q_\lambda(\beta)=\frac{1}{2n}\|y-X\beta\|_2^2
                    +\sum_{j=1}^p p_\lambda(|\beta_j|).
 \label{eq:maincriterion}
\end{equation}
Equation~\eqref{eq:maincriterion} combines squared prediction loss with
coordinatewise penalization. Centering profiles out an unpenalized intercept;
training means and scales are used when predicting new observations. A global
minimizer defines the statistical target, whereas the numerical algorithm
may return a stationary point. The results below distinguish these objects.

\subsection{A design principle for releasing shrinkage}
\label{sec:retentiondesign}

The construction specifies how much of a scalar score is retained as its
magnitude increases. For a fixed $\lambda>0$ and a positive score
$z\geq\lambda$, let $R(z)$ denote a candidate retained value. Consider
absolutely continuous rules $R:[\lambda,\infty)\to[0,\infty)$ satisfying
\begin{equation}
 R(\lambda)=0,\qquad 0\leq R(z)\leq z,\qquad
 R'(z)\leq 1+\frac{R(z)}{z}\quad\text{for almost every }z>\lambda.
 \label{eq:mainretentionbudget}
\end{equation}
Equation~\eqref{eq:mainretentionbudget} requires continuous entry from zero,
prevents amplification beyond the unpenalized score, and limits the growth
of its retained fraction. Indeed, the derivative restriction gives
\[
 \frac{d}{dz}\left\{\frac{R(z)}{z}\right\}\leq\frac{1}{z}.
\]
Thus the retained fraction can increase by at most one unit per unit
increase in the log score. Proportional changes in score are treated alike:
rescaling both $z$ and $\lambda$ leaves this restriction unchanged. The unit
bound fixes the rate of release of shrinkage used in the main method; the
extended family in Supplementary Section~\ref*{sec:construction} allows other rates.

Integrating from $\lambda$ to $z$ bounds $R(z)/z$ by $\log(z/\lambda)$.
Together with the upper bound of one, this identifies the largest feasible
retained fraction. Extending the rule by zero below the entry threshold
and by odd symmetry to negative scores gives
\begin{equation}
 T_\lambda(z)=\begin{cases}
 0,&|z|\leq\lambda,\\
 z\log(|z|/\lambda),&\lambda<|z|<e\lambda,\\
 z,&|z|\geq e\lambda.
 \end{cases}
 \label{eq:mainthreshold}
\end{equation}
Equation~\eqref{eq:mainthreshold} eliminates scores up to $\lambda$, shrinks
intermediate scores, and leaves scores beyond $e\lambda$ unchanged. The
transition is continuous and odd. It attains the derivative bound in
Equation~\eqref{eq:mainretentionbudget} throughout the open transition;
the identity branch is also feasible. Hence it is the pointwise largest
rule in the specified class. Supplementary Section~\ref*{sec:construction} gives the
general-family derivation and recovers the penalty by inversion and
integration. Maximality here concerns retention under the stated constraint;
statistical risk also depends on the signal, noise, and design distribution.

The rule enters with right derivative $T'_\lambda(\lambda+)=1$, matching
the initial slope of soft thresholding. On the positive transition its
slope is $1+\log(z/\lambda)$, so retained coefficients grow increasingly
quickly until shrinkage disappears. This provides a continuous passage
from exclusion to full retention with no additional shape selection.
In regression, $z$ is a partial-residual score that depends on the other
coefficients; the scalar rule therefore does not imply a grouping property
for correlated predictors.

\subsection{Curvature and the exact scalar update}
On $0<t<e\lambda$, let $w=W_0(t/\lambda)\in(0,1)$. Differentiating
Equation~\eqref{eq:mainpenalty} gives
\begin{equation}
 p'_\lambda(t)=\lambda e^w(1-w),\qquad
 p''_\lambda(t)=-\frac{w}{1+w},\qquad
 p'''_\lambda(t)=-\frac{1}{\lambda e^w(1+w)^3}.
 \label{eq:mainderivatives}
\end{equation}
Equation~\eqref{eq:mainderivatives} shows that the penalty is increasing and
strictly concave on the open transition. At the origin its right slope is
$\lambda$ and its right curvature is zero. Its slope joins continuously to
zero at the plateau, but its second derivative is undefined at that join.
The signed penalty $b\mapsto p_\lambda(|b|)$ has a cusp at zero.

\begin{proposition}[Sharp curvature bound and scalar minimization]
\label{prop:mainprox}
The function $b\mapsto p_\lambda(|b|)+b^2/4$ is convex, and $1/2$ is the
smallest weak-convexity constant: no smaller $a$ makes
$b\mapsto p_\lambda(|b|)+ab^2/2$ convex. Consequently, for every $z\in\R$,
$b\mapsto(z-b)^2/2+p_\lambda(|b|)$ is $1/2$-strongly convex and its unique
minimizer is $T_\lambda(z)$.
\end{proposition}
Proposition~\ref{prop:mainprox} is proved in Supplementary Section~\ref*{sec:proof1}.

Figure~\ref{fig:penalties} compares the analytic penalty and scalar update
with LASSO, MCP, and SCAD. Its normalization separates the curve shape from
$\lambda$. The different transition shapes do not by themselves imply a
ranking of prediction or support recovery.
\begin{figure}[!htbp]
\centering\spacingset{1}
\includegraphics[width=0.98\textwidth]{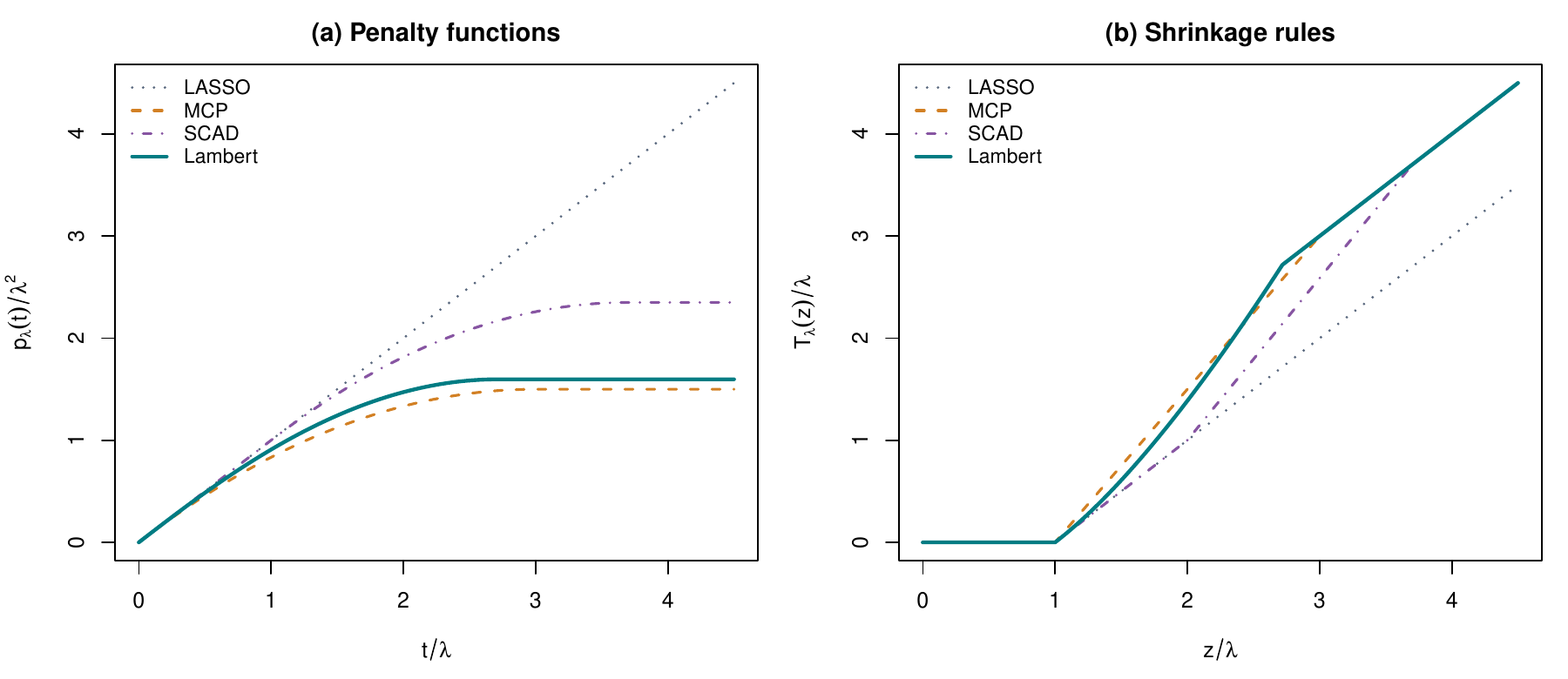}
\caption{Normalized penalty functions (left) and scalar shrinkage rules
(right) for Lambert, LASSO, MCP with $\gamma=3$, and SCAD with $a=3.7$.
Horizontal arguments are divided by $\lambda$; penalties and scalar estimates
are divided by $\lambda^2$ and $\lambda$, respectively. The common entry
threshold is $\lambda$. Lambert, MCP, and SCAD reach identity updates at
$e\lambda$, $3\lambda$, and $3.7\lambda$, respectively, where their penalties
become flat. Only nonnegative arguments are shown.}
\label{fig:penalties}
\end{figure}

\section{Related penalties}
\label{sec:audit}

The relevant distinction is how a penalty controls the transition from
variable entry to full retention. Equation~\eqref{eq:mainretentionbudget}
specifies that choice explicitly for Lambert, and
Equation~\eqref{eq:mainthreshold} is its maximal feasible rule. Neither a
finite plateau nor the use of a special function alone identifies this
construction. To place it in context, we compare its formula and
scalar update with LASSO, SCAD, MCP and its path formulation, logarithmic thresholding,
arctangent regularization, exponential approximations to the counting norm,
transformed-$L_1$ regularization, generic shrinkage-induced penalties, and the
recent minimax logarithmic concave penalty (MLCP)
\citep{fanli2001,zhang2010,mazumderetal2011,malioutovaravkin2014,
selesnickbayram2014,malekmohammadietal2016,zhangxin2018,
chartrand2014,woodworthchartrand2016,liuetal2023,zhangetal2023mlcp}.
At the comparator settings used here, MCP releases shrinkage through a
linear transition, and SCAD uses successive linear pieces. Lambert instead
has a continuously increasing scalar slope on its open logarithmic
transition. Figure~\ref{fig:penalties} shows this difference at a common
entry threshold. Its earlier identity cutoff at these fixed settings
characterizes the scalar rule; after method-specific CV tuning, cutoff
ordering alone cannot rank the fitted models.

The EN-SCAD construction of \citet{biyin2023} is also relevant to the
extended Lambert family. When its
quadratic coefficient is positive, EN-SCAD has
positive curvature near zero, negative curvature in its middle interval, and
an exactly flat tail.  Thus this curvature-sign sequence is not new.  Its
curvature is constant on each open segment and changes at knots, whereas the
Lambert family has strictly declining curvature throughout one analytic
transition. The fixed-shape penalty used in the main study has zero right
curvature at the origin and is strictly concave before its plateau, as
Equation~\eqref{eq:mainderivatives} shows. Positive rescaling cannot make these derivative signatures
identical. This is a distinction between penalty functions; it does not
imply that their multivariate algorithms have the same behavior.

Smooth nonlinear transitions also have direct precedents.
\citet{kulkarni2026smoothscad} construct a smooth SCAD-type rule using a
raised-cosine shrinkage profile. Their rule has an exact zero region and
an identity tail, but differs from Lambert on the intervening interval.
For a positive score $z$ and fixed threshold $\lambda>0$, its scalar slope
rises and then falls back to one at the upper cutoff. In contrast,
Equation~\eqref{eq:mainthreshold} gives Lambert slope
$1+\log(z/\lambda)$ for $\lambda<z<e\lambda$, increasing strictly from one
toward two before returning to one on the identity tail. This derivative
comparison distinguishes the two rules even when their entry and identity
cutoffs are matched. The contribution studied here is therefore the
particular retention constraint and its logarithmic solution, rather than
nonlinearity or continuity of the transition alone.

This structural comparison has a broader scope than the numerical benchmark.
The latter uses LASSO, ridge, and ENET to represent established convex
regularization choices, and MCP and SCAD to represent widely used bounded
nonconvex penalties with reduced shrinkage of large coefficients. This set
allows a focused comparison of prediction, sparsity, and selection with
standard implementations. The additional penalties discussed here clarify
how the Lambert formula differs from existing constructions; their inclusion
in this section does not imply that their empirical performance was tested.
No numerical superiority over those untested methods is claimed.

The comparison considers derivatives where defined, finite versus
asymptotic flattening, threshold cutoffs, and scalar proximal operators.
Positive rescalings of the argument and amplitude are considered to
distinguish a different penalty from a different parameterization.

MLCP provides a particularly relevant comparison because it combines
logarithms with a finite plateau.  For a coefficient magnitude $t\geq0$, its parameters
$\lambda_M>0$, $\gamma>0$, and $\varepsilon>0$, and the notation used here to
avoid confusing its scale with the proposed penalty level, its scalar formula
is \citep{zhangetal2023mlcp}
\begin{equation}
 p^{\mathrm{MLCP}}_{\lambda_M,\gamma,\varepsilon}(t)
 =\begin{cases}
 \displaystyle \lambda_M\log(1+t/\varepsilon)
 -\frac{\log^2(1+t/\varepsilon)}{2\gamma},
 &0\leq t\leq \varepsilon\{e^{\gamma\lambda_M}-1\},\\[6pt]
 \displaystyle \frac{\gamma\lambda_M^2}{2},
 &t>\varepsilon\{e^{\gamma\lambda_M}-1\}.
 \end{cases}
 \label{eq:mlcp}
\end{equation}
Equation~\eqref{eq:mlcp} expresses the MLCP penalty on its increasing
branch and constant tail.  On its first branch,
$(p^{\mathrm{MLCP}})''(0+)=-(\lambda_M+1/\gamma)/\varepsilon^2<0$.
The extended Lambert family in Supplementary Section~\ref*{sec:family} instead has
$p''_{\lambda,c}(0+)=c-1\geq0$.  A transformation
$\widetilde p(t)=A p(Bt)$ with $A,B>0$ multiplies origin curvature by
$AB^2$ and therefore preserves its sign.  Consequently the two penalties
cannot coincide under this rescaling class.  MLCP's authors also motivate an
equivalent surrogate because its direct proximal solution is unavailable,
whereas the Lambert threshold is explicit.

Table~\ref{tab:formula-audit} summarizes these structural differences.  MCP and SCAD have
finite flat tails, but their open pieces are at most quadratic and hence have
zero third derivative.  Logarithmic, arctangent, transformed-$L_1$, and
exponential penalties do not become exactly flat at a finite magnitude.  The
piecewise-exponential penalty studied by \citet{liuetal2023} uses Lambert $W$
in a different proximal calculation; the special function by itself is not a
novelty claim \citep{corlessetal1996}.

\begin{table}[!htbp]
\centering\spacingset{1}
\caption{Curvature and tail behavior of Lambert and related nonconvex sparsity penalties.
The origin column describes right curvature with respect to coefficient
magnitude. The Lambert row covers the extended family; the main penalty
has zero right curvature at the origin.}
\label{tab:formula-audit}
\small

\begin{tabular}{llll}
\toprule
Penalty & Origin & Transition & Tail\\
\midrule
Lambert & Nonnegative & Smoothly declining curvature & Exactly flat\\
EN-SCAD & Positive with quadratic term & Piecewise constant curvature & Exactly flat\\
MCP & Negative & Quadratic piece & Exactly flat\\
SCAD & Zero & Linear then quadratic pieces & Exactly flat\\
MLCP & Negative & Logarithmic concave piece & Exactly flat\\
\bottomrule
\end{tabular}

\end{table}

The penalties in Table~\ref{tab:formula-audit} differ from Lambert in their
threshold rules, derivative structure, or tail behavior. Within these
comparisons, the proposed formula is not a positive argument-and-amplitude
rescaling of an existing penalty. The specific contribution combines a
retention constraint, its logarithmic solution, and a bounded induced
penalty with an explicit coordinate update. The theoretical results concern
this construction under their stated assumptions; the general principles
of shrinkage-induced penalties and folded-concave estimation remain the
foundation for that analysis. Oracle behavior, finite-tail
unbiasedness, correlated-design goals, use of Lambert $W$, and construction
from shrinkage rules have precedents
\citep{zouzhang2009,huangetal2016,zhangzhang2012,fanxuezou2014}; these properties
provide context for the analysis rather than separate novelty claims.
The comparison establishes the stated structural distinctions within the
reviewed class, without excluding an equivalent formulation elsewhere.

\section{Multivariate optimization guarantees}
\label{sec:optimization}
Let $G=X^{\mathsf T}X/n$ be the standardized Gram matrix, let $I_p$ be the
$p$-dimensional identity matrix, and let $\lambda_{\min}(G)$ denote the
smallest eigenvalue of $G$. Scalar uniqueness does not imply multivariate
uniqueness. The following sufficient condition makes that distinction precise.

\begin{theorem}[Sufficient condition for a unique global solution]
\label{thm:globalunique}
If $G-I_p/2$ is positive semidefinite, $Q_\lambda$ is convex. If
$\mu=\lambda_{\min}(G)-1/2>0$, it is $\mu$-strongly convex and coercive,
has a unique global minimizer, and every stationary point equals that minimizer.
\end{theorem}
Theorem~\ref{thm:globalunique} is proved in Supplementary Section~\ref*{sec:proof2}.

Theorem~\ref{thm:globalunique} is sufficient, not necessary, and is not a
claim that the empirical designs satisfy its eigenvalue condition. In
particular, it cannot apply to a rank-deficient design. A bounded penalty
alone does not make a least-squares objective coercive along null directions.

\begin{theorem}[Exact cyclic coordinate descent]
\label{thm:cd}
Fix the data and $\lambda>0$, with unit diagonal entries of $G$. Starting
from a finite vector, update coordinates in the fixed cyclic order
$1,\ldots,p$, using their unique exact minimizers. Denote the iterate after
$k$ individual updates by $b^k$. Then
\begin{equation}
 Q_\lambda(b^k)-Q_\lambda(b^{k+1})
 \geq\frac14\|b^{k+1}-b^k\|_2^2,\qquad
 \sum_{k=0}^{\infty}\|b^{k+1}-b^k\|_2^2<\infty.
 \label{eq:maindescent}
\end{equation}
If the iterates are bounded, every accumulation point is a coordinatewise
minimum and a stationary point of $Q_\lambda$. If its initial sublevel set
contains only finitely many stationary points, the bounded sequence converges
to one of them. Full column rank guarantees boundedness. Under the strict
condition in Theorem~\ref{thm:globalunique}, the sequence converges to the
unique global minimizer.
\end{theorem}

Equation~\eqref{eq:maindescent} quantifies decrease per individual update;
its summability conclusion implies vanishing steps, not an iteration-complexity
bound. Here stationarity means that, with $g=X^{\mathsf T}y/n$, every coordinate
$j$ satisfies $(G b-g)_j+\operatorname{sign}(b_j)p'_\lambda(|b_j|)=0$ when
$b_j\ne0$, and $|(G b-g)_j|\leq\lambda$ when $b_j=0$.
Theorem~\ref{thm:cd} is proved in Supplementary Section~\ref*{sec:proof3}.

Theorem~\ref{thm:cd} concerns exact updates at a fixed penalty level. It does
not certify finite-budget termination, numerical global optimality outside
the strong-convexity regime, or bounded iterates for arbitrary high-dimensional
designs. These are distinct from the verified KKT and curvature checks used
in the computations. Related coordinate-descent convergence and convexity
diagnostics for nonconvex regression are developed by
\citet{brehenyhuang2011,mazumderetal2011}.

\section{Fixed-dimensional oracle equivalence}
\label{sec:mainoracle}
For this asymptotic result let $y_n=X_n\beta^0+\varepsilon_n$, where
$p$ and the vector $\beta^0\in\R^p$ are fixed, $X_n$ is deterministic,
$G_n=X_n^{\mathsf T}X_n/n\to\Sigma\succ0$, and
$\varepsilon_n\sim N(0,\sigma^2I_n)$ with $0<\sigma^2<\infty$.
Here $\succ0$ denotes positive definiteness. Let $S$ be the nonzero support
of $\beta^0$, let $N$ be its complement in $\{1,\ldots,p\}$, and assume
each nonzero signal has fixed positive magnitude. This is a result for the
linear criterion with no penalized intercept; an intercept, when present,
can be profiled out by centering the design and response.

Let $Q_{n,\lambda}$ denote Equation~\eqref{eq:maincriterion} at sample size
$n$. For deterministic positive sequences
$\underline\lambda_n\leq\overline\lambda_n$, assume
\begin{equation}
 \overline\lambda_n\to0,\qquad
 \sqrt n\,\underline\lambda_n\to\infty.
 \label{eq:mainband}
\end{equation}
Equation~\eqref{eq:mainband} makes the upper penalty level vanish on fixed
signals while the lower level dominates the noise score. Define
$\widehat\beta^{\mathrm{or}}_S=(X_{n,S}^{\mathsf T}X_{n,S})^{-1}
X_{n,S}^{\mathsf T}y_n$ and $\widehat\beta^{\mathrm{or}}_N=0$, with
$\widehat\beta^{\mathrm{or}}=0$ if $S$ is empty. This oracle estimator knows
the active support and fits least squares only on that support.

\begin{theorem}[Uniform oracle equivalence for Lambert]
\label{thm:mainoracle}
Under these assumptions, with probability tending to one, every global
minimizer of $Q_{n,\lambda}$ equals $\widehat\beta^{\mathrm{or}}$
simultaneously for all $\lambda\in[\underline\lambda_n,\overline\lambda_n]$.
Thus a possibly data-dependent $\widehat\lambda_n$ lying in this interval
with probability tending to one yields support consistency. If $S$ is
nonempty, the active estimates also satisfy
\begin{equation}
 \sqrt n(\widehat\beta_S-\beta^0_S)
 \ \Rightarrow\ N(0,\sigma^2\Sigma_{SS}^{-1}),
 \label{eq:mainoraclelimit}
\end{equation}
where $\Rightarrow$ denotes convergence in distribution and $\Sigma_{SS}$
is the active principal submatrix of $\Sigma$.
\end{theorem}
Equation~\eqref{eq:mainoraclelimit} gives the same first-order distribution
as least squares with the true support. The statement concerns all global
minimizers, rather than asserting that every stationary point is oracle.
Theorem~\ref{thm:mainoracle} is proved in Supplementary Section~\ref*{sec:proof4}.

Theorem~\ref{thm:mainoracle} does not require the stronger finite-sample
eigenvalue threshold in Theorem~\ref{thm:globalunique}. It is nevertheless
conditional on the asymptotic tuning band: ordinary prediction CV is not
proved to enter that band. It does not cover increasing dimension, vanishing
signals, correlations approaching singularity, or general stationary outputs.

\section{Coordinate updates and tuning}
\label{sec:implementation}

The scalar theorem gives an exact coordinate update after centering the
response and standardizing every training predictor to squared norm $n$.
For coordinate $j$, define the partial residual
$r_j=y_n-\sum_{\ell\neq j}X_{n,\ell}\beta_\ell$ and its score
$z_j=X_{n,j}^{\mathsf T}r_j/n$.  A cyclic step is then
\begin{equation*}
 z_j=\frac{X_{n,j}^{\mathsf T}r_j}{n},
 \qquad
 \beta_j\leftarrow T_\lambda(z_j),
 \qquad j=1,\ldots,p.
\end{equation*}
This normalized update implements the scalar shrinkage rule. Its
validity requires the unit coordinate curvature already assumed in
Equation~\eqref{eq:maincriterion}; within each cross-validation training split,
centering and scaling were estimated from that split alone.

Only the penalty-grid position is selected for Lambert by five-fold
cross-validation. Let $\mathcal I_f$ be
validation fold $f$, let $K=5$, and let $k\in\{0,\ldots,59\}$ index the
common dimensionless grid defined in Equation~\eqref{eq:lambdapath}. Let
$\widehat\beta^{(-f,k)}$ and $\widehat\alpha^{(-f,k)}$ be the coefficients
and unpenalized intercept fitted without that fold, using its training-specific
penalty scale. The selection score is
\begin{equation}
 \operatorname{CV}(k)
 =\frac{1}{K}\sum_{f=1}^{K}
 \frac{1}{|\mathcal I_f|}
 \sum_{i\in\mathcal I_f}
 \left(y_i-\widehat\alpha^{(-f,k)}
 -x_i^{\mathsf T}\widehat\beta^{(-f,k)}\right)^2.
 \label{eq:cvscore}
\end{equation}
Equation~\eqref{eq:cvscore} defines the held-out squared-error criterion;
$x_i\in\R^p$ and $y_i\in\R$ are original-scale predictors and responses.
The chosen position minimizes this score; a relative tolerance
$10^{-10}$ resolves numerical ties in favor of the larger penalty. The final
fit uses the chosen position with the full training sample's entry scale.
The outer test sample and true support are unavailable to this rule.

For LASSO, MCP, SCAD, and Lambert fits, each training split used its own entry
score $\lambda_0=\max_j|Z_j^{\mathsf T}y_c|/m$, where $m$ is the split size,
$Z$ is the centered and standardized design, and $y_c$ is the centered
response.  The common dimensionless path was
\begin{equation}
 \tau_k=10^{-3k/59},
 \qquad
 \lambda_k=\lambda_0\tau_k,
 \qquad k=0,\ldots,59.
 \label{eq:lambdapath}
\end{equation}
Equation~\eqref{eq:lambdapath} defines 60 decreasing levels over three orders
of magnitude. MCP uses $\gamma=3$ and SCAD uses $a=3.7$. These comparator constants match
the defaults of the \texttt{ncvreg} R package (version 3.16.0).
Only the penalty-grid position is selected for these three methods.
Elastic net \citep{zouhastie2005}, abbreviated ENET, combines $L_1$ and
$L_2$ penalties. LASSO, ridge, and ENET were fitted using coordinate-descent
paths \citep{friedmanetal2010}; MCP and SCAD used nonconvex coordinate
descent \citep{brehenyhuang2011}.

At each nonconvex path point, the implementation compared a validated warm
start with a zero start and retained the valid solution with smaller objective
value.  An incomplete solution was not forwarded to the next level.  The
selected fit had to satisfy finite-value checks, a normalized KKT tolerance of
$10^{-7}$, and nonnegative active-set local curvature within the stated
tolerance.  These checks certify the recorded stationary solution; they are
not a global-minimizer certificate for the multivariate nonconvex objective.

\section{Simulation study}
\label{sec:fixedempirical}

\subsection{Design and evaluation}
We compared Lambert, LASSO, ridge, ENET, MCP, and SCAD under the tuning
procedure in Section~\ref{sec:implementation}. The design varies predictor
dimension, correlation, signal strength, and coefficient heterogeneity.
A factorial comparison is supplemented by extreme-signal stress scenarios.
The retention rule in Section~\ref{sec:retentiondesign} provides a framework
for interpreting prediction with strong signals, sensitivity to unequal
active magnitudes, and the effect of correlation on prediction and support
recovery. These comparisons assess the complete CV-tuned procedures;
the oracle result in Theorem~\ref{thm:mainoracle} has fixed-dimensional
asymptotic assumptions and does not establish the finite-sample rankings.

Let $s\in\{1,\ldots,52\}$ index scenarios, $r\in\{1,\ldots,300\}$ index
independent replications within a scenario, and $i$ index observations.
Predictor indices are $j,k\in\{1,\ldots,p_s\}$. The data-generating model is
\begin{equation}
 x_{sr,i}\sim N_{p_s}(0,\Sigma_s),\quad
 (\Sigma_s)_{jk}=\rho_s^{|j-k|},\quad
 y_{sr,i}=x_{sr,i}^{\mathsf T}\beta_s^0+\varepsilon_{sr,i},\quad
 \varepsilon_{sr,i}\sim N(0,\sigma_s^2).
 \label{eq:fixed_dgp}
\end{equation}
Equation~\eqref{eq:fixed_dgp} specifies independent Gaussian observations;
errors are independent of predictors and of one another. Here $x_{sr,i}$ is
a $p_s$-dimensional predictor vector, $\beta_s^0$ is the true coefficient
vector, and the population intercept is zero. For $\rho_s=0$, $\Sigma_s$ is
the identity matrix. Fitted intercepts remain unpenalized.

Each training sample has $n=250$ observations. The factorial design combines
$p_s\in\{125,500\}$, $\rho_s\in\{0,0.3,0.6,0.9\}$, population signal-to-noise ratios
$\mathrm{SNR}_s\in\{0.25,1,4\}$, and two signal profiles. The first ten
coefficients are active, with five positive signs followed by five negative
signs. The equal profile assigns all ten the same magnitude. The
moderately heterogeneous profile uses magnitudes $3,3,2,2,2,2,1,1,1,1$ in
that order. Four additional stress scenarios combine both dimensions with
$\rho\in\{0,0.9\}$ and SNR 1, using extreme magnitudes
$4,3,2,1.5,1,0.7,0.5,0.3,0.2,0.1$. Each vector is normalized to Euclidean norm one. The noise
variance is calibrated by
\begin{equation}
 \sigma_s^2=\frac{(\beta_s^0)^{\mathsf T}\Sigma_s\beta_s^0}{\mathrm{SNR}_s}.
 \label{eq:fixed_variance}
\end{equation}
Under Equation~\eqref{eq:fixed_dgp}, the numerator in
Equation~\eqref{eq:fixed_variance} is the population variance of the linear
signal. This calibration fixes the signal-to-noise ratio across covariance
and magnitude profiles. Each replication has an independent test sample
of $n_{\rm test}=2000$ observations. All methods share
the training sample, test sample, and five training-CV folds within a
replication. The two predictor dimensions examine both $p<n$ and $p>n$.

The penalty shapes and regularization path follow
Section~\ref{sec:implementation}. Only the penalty level is tuned for
Lambert, MCP, and SCAD; ENET additionally uses the mixing grid
$\{0,0.25,0.5,0.75,1\}$. Neither true support nor test observations enter
tuning. Early stopping is disabled, and selected coefficients are neither
removed nor refitted after estimation.

For method $m$, let $\widehat y_{msr,i}$ be its prediction on independent
test observation $i$. We evaluate normalized test mean squared error as
\begin{equation}
 \mathrm{NMSE}_{msr}=\frac{1}{n_{\rm test}\sigma_s^2}
 \sum_{i=1}^{n_{\rm test}}
 (y_{sr,i}^{\rm test}-\widehat y_{msr,i})^2.
 \label{eq:fixed_nmse}
\end{equation}
Equation~\eqref{eq:fixed_nmse} divides test MSE by population error variance;
lower values are preferable and a value near one approaches the irreducible
noise level in expectation. We also assess coefficient error (CE) and
normalized excess prediction risk (ER). With
$d_{msr}=\widehat\beta_{msr}-\beta_s^0$ on the original coefficient scale
and fitted intercept $\widehat\alpha_{msr}$, these are
\begin{equation}
 \mathrm{CE}_{msr}=\|d_{msr}\|_2^2,\qquad
 \mathrm{ER}_{msr}=
 \frac{\widehat\alpha_{msr}^{\,2}+d_{msr}^{\mathsf T}\Sigma_s d_{msr}}
 {\sigma_s^2}.
 \label{eq:fixed_secondary}
\end{equation}
Equation~\eqref{eq:fixed_secondary} uses the zero predictor means and zero
true intercept in Equation~\eqref{eq:fixed_dgp}. ER excludes irreducible
test noise and describes conditional population prediction risk, whereas CE
measures coefficient accuracy without covariance weighting.

Support reporting counts an original-scale coefficient as selected when its
absolute value exceeds $10^{-8}$; this tolerance does not alter predictions.
Let TP, FN, FP, and TN count selected active, missed active, selected
inactive, and excluded inactive predictors, respectively. For positive
denominator, support agreement is measured by
\begin{equation}
 \mathrm{MCC}=
 \frac{\mathrm{TP}\,\mathrm{TN}-\mathrm{FP}\,\mathrm{FN}}
 {\sqrt{(\mathrm{TP}+\mathrm{FP})(\mathrm{TP}+\mathrm{FN})
 (\mathrm{TN}+\mathrm{FP})(\mathrm{TN}+\mathrm{FN})}}.
 \label{eq:fixedmcc}
\end{equation}
Equation~\eqref{eq:fixedmcc} is calculated separately for each replication
using the \texttt{mcc} function in the \texttt{mltools} R package
(version 0.3.5); larger values indicate better agreement.
The function returns zero when the denominator is zero. Ridge is included
as a prediction comparator; its essentially dense fits do not provide
meaningful variable selection.

These metrics address different uses. NMSE evaluates future-response
accuracy relative to noise, permitting comparison across error variances.
MCC addresses variable discovery, balancing false inclusions and omissions
against the known support. CE assesses coefficient reconstruction when
regression effects themselves are of interest. ER isolates prediction loss
from estimation, excluding irreducible noise and test-sample variation.
Disagreement among these criteria reveals whether predictive gains also
improve support or coefficient recovery.

Results are summarized by arithmetic means and Monte Carlo standard
errors (MCSEs) across replications. For a replication-level score $v_r$, including a
paired Lambert-minus-comparator difference on the same sample, the MCSE is
\begin{equation}
 \mathrm{MCSE}(\bar v)=\frac{\operatorname{sd}(v_1,\ldots,v_B)}{\sqrt B},
 \qquad B=300.
 \label{eq:fixed_mcse}
\end{equation}
Equation~\eqref{eq:fixed_mcse} quantifies Monte Carlo uncertainty in the
scenario mean using the independent replications. Paired 95\% intervals use
the Student $t_{299}$ critical value and are descriptive, without multiplicity
adjustment. They do not establish simultaneous superiority. Main-design averages weight its 48 scenarios equally, excluding the
stress scenarios. Scenario-specific results remain the primary basis for
comparison; the main-design mixture has no pooled MCSE.

\subsection{Prediction, selection, and coefficient estimation}
Tables~\ref{tab:nmse_combined}--\ref{tab:supp_excess_risk} report NMSE, MCC,
CE, and ER, respectively. In these tables, E, H, and X denote equal,
moderately heterogeneous, and extreme signal profiles; $\rho$ is the AR(1)
correlation. Parentheses contain MCSEs. Bold identifies the best unrounded
value within each scenario, including all exact ties: smaller values for
NMSE, CE, and ER, and larger values for MCC.

\begin{table}[!tbp]
\captionsetup{aboveskip=0pt,belowskip=3pt}
\centering\spacingset{1}\scriptsize
\setlength{\tabcolsep}{2pt}
\renewcommand{\arraystretch}{0.90}
\caption{Out-of-sample prediction error across signal profiles and predictor correlations: normalized mean squared error (NMSE).}
\label{tab:nmse_combined}
\resizebox{\textwidth}{!}{
\begin{tabular}{>{\centering\arraybackslash}m{0.55cm}>{\centering\arraybackslash}m{0.5cm}>{\centering\arraybackslash}m{0.9cm}>{\centering\arraybackslash}m{0.7cm}>{\raggedleft\arraybackslash}m{2.28cm}>{\raggedleft\arraybackslash}m{2.28cm}>{\raggedleft\arraybackslash}m{2.28cm}>{\raggedleft\arraybackslash}m{2.28cm}>{\raggedleft\arraybackslash}m{2.28cm}>{\raggedleft\arraybackslash}m{2.28cm}}
\toprule
$p$ & $\rho$ & Signal & SNR & Lambert & LASSO & Ridge & ENET & MCP & SCAD\\
\midrule
125 & 0.0 & E & 0.25 & \mbox{1.1812\,{\scriptsize (0.0033)}} & \textbf{\mbox{1.1631\,{\scriptsize (0.0032)}}} & \mbox{1.1903\,{\scriptsize (0.0024)}} & \mbox{1.1635\,{\scriptsize (0.0029)}} & \mbox{1.1807\,{\scriptsize (0.0033)}} & \mbox{1.1693\,{\scriptsize (0.0032)}}\\

 &  &  & 1.00 & \textbf{\mbox{1.1147\,{\scriptsize (0.0041)}}} & \mbox{1.1817\,{\scriptsize (0.0040)}} & \mbox{1.4395\,{\scriptsize (0.0042)}} & \mbox{1.1829\,{\scriptsize (0.0041)}} & \mbox{1.1236\,{\scriptsize (0.0040)}} & \mbox{1.1309\,{\scriptsize (0.0041)}}\\

 &  &  & 4.00 & \textbf{\mbox{1.0455\,{\scriptsize (0.0024)}}} & \mbox{1.1720\,{\scriptsize (0.0039)}} & \mbox{1.7268\,{\scriptsize (0.0074)}} & \mbox{1.1720\,{\scriptsize (0.0039)}} & \mbox{1.0478\,{\scriptsize (0.0025)}} & \mbox{1.0482\,{\scriptsize (0.0024)}}\\

 &  & H & 0.25 & \mbox{1.1538\,{\scriptsize (0.0033)}} & \textbf{\mbox{1.1455\,{\scriptsize (0.0031)}}} & \mbox{1.1943\,{\scriptsize (0.0026)}} & \mbox{1.1499\,{\scriptsize (0.0031)}} & \mbox{1.1525\,{\scriptsize (0.0034)}} & \mbox{1.1468\,{\scriptsize (0.0030)}}\\

 &  &  & 1.00 & \textbf{\mbox{1.1347\,{\scriptsize (0.0037)}}} & \mbox{1.1711\,{\scriptsize (0.0037)}} & \mbox{1.4335\,{\scriptsize (0.0041)}} & \mbox{1.1715\,{\scriptsize (0.0037)}} & \mbox{1.1375\,{\scriptsize (0.0037)}} & \mbox{1.1371\,{\scriptsize (0.0036)}}\\

 &  &  & 4.00 & \textbf{\mbox{1.0831\,{\scriptsize (0.0030)}}} & \mbox{1.1749\,{\scriptsize (0.0039)}} & \mbox{1.7303\,{\scriptsize (0.0068)}} & \mbox{1.1749\,{\scriptsize (0.0039)}} & \mbox{1.0864\,{\scriptsize (0.0030)}} & \mbox{1.0940\,{\scriptsize (0.0030)}}\\

 &  & X & 1.00 & \textbf{\mbox{1.0989\,{\scriptsize (0.0030)}}} & \mbox{1.1290\,{\scriptsize (0.0032)}} & \mbox{1.4345\,{\scriptsize (0.0042)}} & \mbox{1.1291\,{\scriptsize (0.0032)}} & \mbox{1.0996\,{\scriptsize (0.0030)}} & \mbox{1.0995\,{\scriptsize (0.0028)}}\\

 & 0.3 & E & 0.25 & \mbox{1.1434\,{\scriptsize (0.0030)}} & \textbf{\mbox{1.1254\,{\scriptsize (0.0028)}}} & \mbox{1.1559\,{\scriptsize (0.0025)}} & \mbox{1.1279\,{\scriptsize (0.0028)}} & \mbox{1.1447\,{\scriptsize (0.0030)}} & \mbox{1.1327\,{\scriptsize (0.0029)}}\\

 &  &  & 1.00 & \mbox{1.1787\,{\scriptsize (0.0048)}} & \textbf{\mbox{1.1495\,{\scriptsize (0.0035)}}} & \mbox{1.3261\,{\scriptsize (0.0038)}} & \mbox{1.1530\,{\scriptsize (0.0037)}} & \mbox{1.1756\,{\scriptsize (0.0045)}} & \mbox{1.1830\,{\scriptsize (0.0041)}}\\

 &  &  & 4.00 & \textbf{\mbox{1.0561\,{\scriptsize (0.0025)}}} & \mbox{1.1517\,{\scriptsize (0.0038)}} & \mbox{1.5676\,{\scriptsize (0.0060)}} & \mbox{1.1517\,{\scriptsize (0.0038)}} & \mbox{1.0617\,{\scriptsize (0.0026)}} & \mbox{1.0644\,{\scriptsize (0.0026)}}\\

 &  & H & 0.25 & \mbox{1.1278\,{\scriptsize (0.0027)}} & \textbf{\mbox{1.1129\,{\scriptsize (0.0026)}}} & \mbox{1.1588\,{\scriptsize (0.0024)}} & \mbox{1.1172\,{\scriptsize (0.0027)}} & \mbox{1.1264\,{\scriptsize (0.0027)}} & \mbox{1.1177\,{\scriptsize (0.0026)}}\\

 &  &  & 1.00 & \mbox{1.1513\,{\scriptsize (0.0039)}} & \textbf{\mbox{1.1429\,{\scriptsize (0.0034)}}} & \mbox{1.3337\,{\scriptsize (0.0038)}} & \mbox{1.1445\,{\scriptsize (0.0034)}} & \mbox{1.1483\,{\scriptsize (0.0039)}} & \mbox{1.1494\,{\scriptsize (0.0037)}}\\

 &  &  & 4.00 & \mbox{1.1138\,{\scriptsize (0.0035)}} & \mbox{1.1562\,{\scriptsize (0.0037)}} & \mbox{1.5839\,{\scriptsize (0.0059)}} & \mbox{1.1562\,{\scriptsize (0.0037)}} & \textbf{\mbox{1.1097\,{\scriptsize (0.0035)}}} & \mbox{1.1138\,{\scriptsize (0.0033)}}\\

 & 0.6 & E & 0.25 & \mbox{1.1161\,{\scriptsize (0.0026)}} & \textbf{\mbox{1.0942\,{\scriptsize (0.0026)}}} & \mbox{1.1303\,{\scriptsize (0.0025)}} & \mbox{1.0965\,{\scriptsize (0.0026)}} & \mbox{1.1134\,{\scriptsize (0.0027)}} & \mbox{1.1024\,{\scriptsize (0.0025)}}\\

 &  &  & 1.00 & \mbox{1.1974\,{\scriptsize (0.0039)}} & \textbf{\mbox{1.1314\,{\scriptsize (0.0030)}}} & \mbox{1.2412\,{\scriptsize (0.0032)}} & \mbox{1.1335\,{\scriptsize (0.0031)}} & \mbox{1.2003\,{\scriptsize (0.0037)}} & \mbox{1.1976\,{\scriptsize (0.0033)}}\\

 &  &  & 4.00 & \textbf{\mbox{1.1168\,{\scriptsize (0.0049)}}} & \mbox{1.1460\,{\scriptsize (0.0033)}} & \mbox{1.4051\,{\scriptsize (0.0043)}} & \mbox{1.1466\,{\scriptsize (0.0033)}} & \mbox{1.1238\,{\scriptsize (0.0050)}} & \mbox{1.1220\,{\scriptsize (0.0048)}}\\

 &  & H & 0.25 & \mbox{1.1095\,{\scriptsize (0.0027)}} & \textbf{\mbox{1.0884\,{\scriptsize (0.0026)}}} & \mbox{1.1319\,{\scriptsize (0.0025)}} & \mbox{1.0908\,{\scriptsize (0.0026)}} & \mbox{1.1065\,{\scriptsize (0.0026)}} & \mbox{1.1003\,{\scriptsize (0.0026)}}\\

 &  &  & 1.00 & \mbox{1.1728\,{\scriptsize (0.0039)}} & \textbf{\mbox{1.1232\,{\scriptsize (0.0029)}}} & \mbox{1.2370\,{\scriptsize (0.0034)}} & \mbox{1.1257\,{\scriptsize (0.0030)}} & \mbox{1.1731\,{\scriptsize (0.0038)}} & \mbox{1.1701\,{\scriptsize (0.0035)}}\\

 &  &  & 4.00 & \textbf{\mbox{1.1387\,{\scriptsize (0.0044)}}} & \mbox{1.1446\,{\scriptsize (0.0036)}} & \mbox{1.4014\,{\scriptsize (0.0045)}} & \mbox{1.1452\,{\scriptsize (0.0037)}} & \mbox{1.1438\,{\scriptsize (0.0043)}} & \mbox{1.1420\,{\scriptsize (0.0043)}}\\

 & 0.9 & E & 0.25 & \mbox{1.0843\,{\scriptsize (0.0030)}} & \textbf{\mbox{1.0828\,{\scriptsize (0.0026)}}} & \mbox{1.1190\,{\scriptsize (0.0027)}} & \mbox{1.0863\,{\scriptsize (0.0027)}} & \mbox{1.0893\,{\scriptsize (0.0031)}} & \mbox{1.0844\,{\scriptsize (0.0027)}}\\

 &  &  & 1.00 & \mbox{1.1472\,{\scriptsize (0.0038)}} & \textbf{\mbox{1.1094\,{\scriptsize (0.0027)}}} & \mbox{1.1790\,{\scriptsize (0.0030)}} & \mbox{1.1108\,{\scriptsize (0.0029)}} & \mbox{1.1512\,{\scriptsize (0.0034)}} & \mbox{1.1509\,{\scriptsize (0.0035)}}\\

 &  &  & 4.00 & \mbox{1.1747\,{\scriptsize (0.0049)}} & \textbf{\mbox{1.1421\,{\scriptsize (0.0032)}}} & \mbox{1.2758\,{\scriptsize (0.0038)}} & \mbox{1.1436\,{\scriptsize (0.0032)}} & \mbox{1.2022\,{\scriptsize (0.0049)}} & \mbox{1.1920\,{\scriptsize (0.0045)}}\\

 &  & H & 0.25 & \mbox{1.0716\,{\scriptsize (0.0028)}} & \textbf{\mbox{1.0703\,{\scriptsize (0.0024)}}} & \mbox{1.1042\,{\scriptsize (0.0026)}} & \mbox{1.0720\,{\scriptsize (0.0024)}} & \mbox{1.0730\,{\scriptsize (0.0028)}} & \mbox{1.0726\,{\scriptsize (0.0028)}}\\

 &  &  & 1.00 & \mbox{1.1194\,{\scriptsize (0.0033)}} & \textbf{\mbox{1.0998\,{\scriptsize (0.0028)}}} & \mbox{1.1607\,{\scriptsize (0.0031)}} & \mbox{1.1029\,{\scriptsize (0.0030)}} & \mbox{1.1187\,{\scriptsize (0.0033)}} & \mbox{1.1185\,{\scriptsize (0.0033)}}\\

 &  &  & 4.00 & \mbox{1.1498\,{\scriptsize (0.0040)}} & \textbf{\mbox{1.1326\,{\scriptsize (0.0029)}}} & \mbox{1.2397\,{\scriptsize (0.0033)}} & \mbox{1.1343\,{\scriptsize (0.0030)}} & \mbox{1.1672\,{\scriptsize (0.0046)}} & \mbox{1.1613\,{\scriptsize (0.0043)}}\\

 &  & X & 1.00 & \mbox{1.0714\,{\scriptsize (0.0025)}} & \textbf{\mbox{1.0523\,{\scriptsize (0.0023)}}} & \mbox{1.1341\,{\scriptsize (0.0028)}} & \mbox{1.0541\,{\scriptsize (0.0024)}} & \mbox{1.0728\,{\scriptsize (0.0026)}} & \mbox{1.0738\,{\scriptsize (0.0025)}}\\
\cmidrule{1-10}
500 & 0.0 & E & 0.25 & \mbox{1.2174\,{\scriptsize (0.0029)}} & \mbox{1.2147\,{\scriptsize (0.0031)}} & \mbox{1.2448\,{\scriptsize (0.0023)}} & \mbox{1.2172\,{\scriptsize (0.0028)}} & \mbox{1.2205\,{\scriptsize (0.0030)}} & \textbf{\mbox{1.2137\,{\scriptsize (0.0029)}}}\\

 &  &  & 1.00 & \textbf{\mbox{1.1796\,{\scriptsize (0.0055)}}} & \mbox{1.2746\,{\scriptsize (0.0053)}} & \mbox{1.8079\,{\scriptsize (0.0044)}} & \mbox{1.2753\,{\scriptsize (0.0053)}} & \mbox{1.1840\,{\scriptsize (0.0053)}} & \mbox{1.2114\,{\scriptsize (0.0054)}}\\

 &  &  & 4.00 & \textbf{\mbox{1.0514\,{\scriptsize (0.0023)}}} & \mbox{1.2779\,{\scriptsize (0.0054)}} & \mbox{3.6394\,{\scriptsize (0.0109)}} & \mbox{1.2779\,{\scriptsize (0.0054)}} & \mbox{1.0563\,{\scriptsize (0.0024)}} & \mbox{1.0615\,{\scriptsize (0.0025)}}\\

 &  & H & 0.25 & \textbf{\mbox{1.1800\,{\scriptsize (0.0034)}}} & \mbox{1.1827\,{\scriptsize (0.0033)}} & \mbox{1.2436\,{\scriptsize (0.0024)}} & \mbox{1.1876\,{\scriptsize (0.0034)}} & \mbox{1.1814\,{\scriptsize (0.0034)}} & \mbox{1.1808\,{\scriptsize (0.0033)}}\\

 &  &  & 1.00 & \textbf{\mbox{1.1947\,{\scriptsize (0.0049)}}} & \mbox{1.2652\,{\scriptsize (0.0047)}} & \mbox{1.8131\,{\scriptsize (0.0041)}} & \mbox{1.2652\,{\scriptsize (0.0047)}} & \mbox{1.1965\,{\scriptsize (0.0048)}} & \mbox{1.2091\,{\scriptsize (0.0048)}}\\

 &  &  & 4.00 & \mbox{1.1241\,{\scriptsize (0.0037)}} & \mbox{1.2863\,{\scriptsize (0.0048)}} & \mbox{3.6379\,{\scriptsize (0.0103)}} & \mbox{1.2863\,{\scriptsize (0.0048)}} & \textbf{\mbox{1.1223\,{\scriptsize (0.0038)}}} & \mbox{1.1405\,{\scriptsize (0.0036)}}\\

 &  & X & 1.00 & \textbf{\mbox{1.1247\,{\scriptsize (0.0035)}}} & \mbox{1.1829\,{\scriptsize (0.0039)}} & \mbox{1.8107\,{\scriptsize (0.0042)}} & \mbox{1.1829\,{\scriptsize (0.0039)}} & \mbox{1.1252\,{\scriptsize (0.0037)}} & \mbox{1.1324\,{\scriptsize (0.0035)}}\\

 & 0.3 & E & 0.25 & \mbox{1.1715\,{\scriptsize (0.0029)}} & \mbox{1.1667\,{\scriptsize (0.0029)}} & \mbox{1.2251\,{\scriptsize (0.0024)}} & \mbox{1.1708\,{\scriptsize (0.0030)}} & \mbox{1.1748\,{\scriptsize (0.0029)}} & \textbf{\mbox{1.1662\,{\scriptsize (0.0029)}}}\\

 &  &  & 1.00 & \mbox{1.2674\,{\scriptsize (0.0044)}} & \textbf{\mbox{1.2235\,{\scriptsize (0.0040)}}} & \mbox{1.6767\,{\scriptsize (0.0041)}} & \mbox{1.2257\,{\scriptsize (0.0041)}} & \mbox{1.2479\,{\scriptsize (0.0045)}} & \mbox{1.2607\,{\scriptsize (0.0042)}}\\

 &  &  & 4.00 & \textbf{\mbox{1.0733\,{\scriptsize (0.0028)}}} & \mbox{1.2280\,{\scriptsize (0.0046)}} & \mbox{2.9491\,{\scriptsize (0.0096)}} & \mbox{1.2280\,{\scriptsize (0.0046)}} & \mbox{1.0848\,{\scriptsize (0.0030)}} & \mbox{1.0930\,{\scriptsize (0.0029)}}\\

 &  & H & 0.25 & \mbox{1.1465\,{\scriptsize (0.0030)}} & \mbox{1.1460\,{\scriptsize (0.0031)}} & \mbox{1.2272\,{\scriptsize (0.0025)}} & \mbox{1.1508\,{\scriptsize (0.0033)}} & \mbox{1.1476\,{\scriptsize (0.0031)}} & \textbf{\mbox{1.1446\,{\scriptsize (0.0030)}}}\\

 &  &  & 1.00 & \mbox{1.2198\,{\scriptsize (0.0044)}} & \textbf{\mbox{1.2141\,{\scriptsize (0.0039)}}} & \mbox{1.6845\,{\scriptsize (0.0040)}} & \mbox{1.2151\,{\scriptsize (0.0039)}} & \mbox{1.2159\,{\scriptsize (0.0044)}} & \mbox{1.2217\,{\scriptsize (0.0041)}}\\

 &  &  & 4.00 & \mbox{1.1414\,{\scriptsize (0.0037)}} & \mbox{1.2249\,{\scriptsize (0.0045)}} & \mbox{2.9718\,{\scriptsize (0.0105)}} & \mbox{1.2249\,{\scriptsize (0.0045)}} & \textbf{\mbox{1.1343\,{\scriptsize (0.0039)}}} & \mbox{1.1444\,{\scriptsize (0.0036)}}\\

 & 0.6 & E & 0.25 & \mbox{1.1263\,{\scriptsize (0.0027)}} & \textbf{\mbox{1.1216\,{\scriptsize (0.0028)}}} & \mbox{1.2126\,{\scriptsize (0.0024)}} & \mbox{1.1255\,{\scriptsize (0.0029)}} & \mbox{1.1271\,{\scriptsize (0.0029)}} & \mbox{1.1224\,{\scriptsize (0.0028)}}\\

 &  &  & 1.00 & \mbox{1.2432\,{\scriptsize (0.0042)}} & \textbf{\mbox{1.1771\,{\scriptsize (0.0035)}}} & \mbox{1.5755\,{\scriptsize (0.0040)}} & \mbox{1.1797\,{\scriptsize (0.0035)}} & \mbox{1.2435\,{\scriptsize (0.0041)}} & \mbox{1.2495\,{\scriptsize (0.0038)}}\\

 &  &  & 4.00 & \textbf{\mbox{1.2020\,{\scriptsize (0.0078)}}} & \mbox{1.2101\,{\scriptsize (0.0045)}} & \mbox{2.3961\,{\scriptsize (0.0091)}} & \mbox{1.2108\,{\scriptsize (0.0045)}} & \mbox{1.2325\,{\scriptsize (0.0077)}} & \mbox{1.2303\,{\scriptsize (0.0073)}}\\

 &  & H & 0.25 & \mbox{1.1177\,{\scriptsize (0.0027)}} & \textbf{\mbox{1.1098\,{\scriptsize (0.0029)}}} & \mbox{1.2150\,{\scriptsize (0.0025)}} & \mbox{1.1136\,{\scriptsize (0.0031)}} & \mbox{1.1172\,{\scriptsize (0.0027)}} & \mbox{1.1119\,{\scriptsize (0.0027)}}\\

 &  &  & 1.00 & \mbox{1.2146\,{\scriptsize (0.0038)}} & \textbf{\mbox{1.1673\,{\scriptsize (0.0034)}}} & \mbox{1.5759\,{\scriptsize (0.0040)}} & \mbox{1.1694\,{\scriptsize (0.0034)}} & \mbox{1.2135\,{\scriptsize (0.0039)}} & \mbox{1.2202\,{\scriptsize (0.0037)}}\\

 &  &  & 4.00 & \mbox{1.2133\,{\scriptsize (0.0053)}} & \textbf{\mbox{1.2091\,{\scriptsize (0.0037)}}} & \mbox{2.3673\,{\scriptsize (0.0081)}} & \mbox{1.2095\,{\scriptsize (0.0037)}} & \mbox{1.2292\,{\scriptsize (0.0053)}} & \mbox{1.2344\,{\scriptsize (0.0049)}}\\

 & 0.9 & E & 0.25 & \textbf{\mbox{1.0980\,{\scriptsize (0.0034)}}} & \mbox{1.1278\,{\scriptsize (0.0033)}} & \mbox{1.2485\,{\scriptsize (0.0024)}} & \mbox{1.1305\,{\scriptsize (0.0033)}} & \mbox{1.1020\,{\scriptsize (0.0032)}} & \mbox{1.1077\,{\scriptsize (0.0033)}}\\

 &  &  & 1.00 & \mbox{1.1811\,{\scriptsize (0.0037)}} & \textbf{\mbox{1.1591\,{\scriptsize (0.0034)}}} & \mbox{1.5182\,{\scriptsize (0.0048)}} & \mbox{1.1621\,{\scriptsize (0.0035)}} & \mbox{1.1899\,{\scriptsize (0.0037)}} & \mbox{1.1890\,{\scriptsize (0.0034)}}\\

 &  &  & 4.00 & \mbox{1.2551\,{\scriptsize (0.0051)}} & \textbf{\mbox{1.2139\,{\scriptsize (0.0039)}}} & \mbox{1.9779\,{\scriptsize (0.0082)}} & \mbox{1.2157\,{\scriptsize (0.0040)}} & \mbox{1.2838\,{\scriptsize (0.0052)}} & \mbox{1.2744\,{\scriptsize (0.0048)}}\\

 &  & H & 0.25 & \textbf{\mbox{1.0739\,{\scriptsize (0.0024)}}} & \mbox{1.0863\,{\scriptsize (0.0025)}} & \mbox{1.2168\,{\scriptsize (0.0028)}} & \mbox{1.0881\,{\scriptsize (0.0027)}} & \mbox{1.0781\,{\scriptsize (0.0027)}} & \mbox{1.0791\,{\scriptsize (0.0025)}}\\

 &  &  & 1.00 & \textbf{\mbox{1.1374\,{\scriptsize (0.0034)}}} & \mbox{1.1396\,{\scriptsize (0.0031)}} & \mbox{1.4450\,{\scriptsize (0.0042)}} & \mbox{1.1429\,{\scriptsize (0.0032)}} & \mbox{1.1405\,{\scriptsize (0.0035)}} & \mbox{1.1479\,{\scriptsize (0.0035)}}\\

 &  &  & 4.00 & \mbox{1.2129\,{\scriptsize (0.0050)}} & \textbf{\mbox{1.1974\,{\scriptsize (0.0037)}}} & \mbox{1.8098\,{\scriptsize (0.0069)}} & \mbox{1.1989\,{\scriptsize (0.0037)}} & \mbox{1.2393\,{\scriptsize (0.0048)}} & \mbox{1.2354\,{\scriptsize (0.0050)}}\\

 &  & X & 1.00 & \mbox{1.0836\,{\scriptsize (0.0025)}} & \textbf{\mbox{1.0652\,{\scriptsize (0.0027)}}} & \mbox{1.3627\,{\scriptsize (0.0038)}} & \mbox{1.0674\,{\scriptsize (0.0027)}} & \mbox{1.0896\,{\scriptsize (0.0029)}} & \mbox{1.0894\,{\scriptsize (0.0026)}}\\
\bottomrule
\end{tabular}
}
\end{table}

Table~\ref{tab:nmse_combined} favors Lambert most clearly for strong equal
signals, where it leads in six of eight settings. At $p=500$, $\rho=0$,
and SNR 4, NMSE is 1.0514 for Lambert, 1.0563 for MCP, 1.0615 for SCAD,
1.2779 for both LASSO and ENET, and 3.6394 for ridge. The Lambert advantage
over MCP and SCAD extends to $\rho=0.6$. At $\rho=0.9$, however, LASSO
improves on Lambert (1.2139 versus 1.2551). Thus moderate correlation does
not remove Lambert's advantage, but strong correlation can favor a convex
sparse estimator.

Paired differences help distinguish the size of these gains from Monte
Carlo variation. For equal signals at $p=500$, $\rho=0.6$, and SNR 4,
Lambert-minus-MCP NMSE is $-0.0305$ (MCSE 0.0046), and
Lambert-minus-SCAD NMSE is $-0.0283$ (MCSE 0.0042). Both descriptive paired
95\% intervals exclude zero. These are scenario-specific comparisons under
the uncertainty convention in Equation~\eqref{eq:fixed_mcse}.

Across the main design, Lambert has lower NMSE than MCP and SCAD in 33
scenarios each, and lower NMSE than ridge throughout. LASSO instead leads
in ten of the sixteen low-SNR settings. Heterogeneity also changes the
ranking: at $p=500$, $\rho=0$, and SNR 4, MCP slightly improves on Lambert
for the heterogeneous profile (1.1223 versus 1.1241). Lambert improves on both MCP
and SCAD in every extreme-signal scenario, whereas LASSO and ENET are
better at $\rho=0.9$.

\begin{table}[!tbp]
\captionsetup{aboveskip=0pt,belowskip=3pt}
\centering\spacingset{1}\scriptsize
\setlength{\tabcolsep}{2pt}
\renewcommand{\arraystretch}{0.90}
\caption{Variable-selection accuracy across signal profiles and predictor correlations: Matthews correlation coefficient (MCC).}
\label{tab:mcc_combined}
\resizebox{\textwidth}{!}{
\begin{tabular}{>{\centering\arraybackslash}m{0.55cm}>{\centering\arraybackslash}m{0.5cm}>{\centering\arraybackslash}m{0.9cm}>{\centering\arraybackslash}m{0.7cm}>{\raggedleft\arraybackslash}m{2.28cm}>{\raggedleft\arraybackslash}m{2.28cm}>{\raggedleft\arraybackslash}m{2.28cm}>{\raggedleft\arraybackslash}m{2.28cm}>{\raggedleft\arraybackslash}m{2.28cm}>{\raggedleft\arraybackslash}m{2.28cm}}
\toprule
$p$ & $\rho$ & Signal & SNR & Lambert & LASSO & Ridge & ENET & MCP & SCAD\\
\midrule
125 & 0.0 & E & 0.25 & \textbf{\mbox{0.4950\,{\scriptsize (0.0103)}}} & \mbox{0.4455\,{\scriptsize (0.0086)}} & \mbox{0.0000\,{\scriptsize (0.0000)}} & \mbox{0.3457\,{\scriptsize (0.0116)}} & \mbox{0.4922\,{\scriptsize (0.0110)}} & \mbox{0.4755\,{\scriptsize (0.0088)}}\\

 &  &  & 1.00 & \mbox{0.6686\,{\scriptsize (0.0049)}} & \mbox{0.5022\,{\scriptsize (0.0055)}} & \mbox{0.0000\,{\scriptsize (0.0000)}} & \mbox{0.5006\,{\scriptsize (0.0056)}} & \textbf{\mbox{0.7260\,{\scriptsize (0.0056)}}} & \mbox{0.5585\,{\scriptsize (0.0044)}}\\

 &  &  & 4.00 & \mbox{0.9233\,{\scriptsize (0.0052)}} & \mbox{0.5067\,{\scriptsize (0.0051)}} & \mbox{0.0000\,{\scriptsize (0.0000)}} & \mbox{0.5067\,{\scriptsize (0.0051)}} & \textbf{\mbox{0.9272\,{\scriptsize (0.0048)}}} & \mbox{0.8153\,{\scriptsize (0.0060)}}\\

 &  & H & 0.25 & \mbox{0.4470\,{\scriptsize (0.0077)}} & \mbox{0.4006\,{\scriptsize (0.0069)}} & \mbox{0.0000\,{\scriptsize (0.0000)}} & \mbox{0.3444\,{\scriptsize (0.0097)}} & \textbf{\mbox{0.4501\,{\scriptsize (0.0081)}}} & \mbox{0.4235\,{\scriptsize (0.0073)}}\\

 &  &  & 1.00 & \mbox{0.6648\,{\scriptsize (0.0057)}} & \mbox{0.5099\,{\scriptsize (0.0058)}} & \mbox{0.0000\,{\scriptsize (0.0000)}} & \mbox{0.5093\,{\scriptsize (0.0059)}} & \textbf{\mbox{0.6956\,{\scriptsize (0.0058)}}} & \mbox{0.5775\,{\scriptsize (0.0054)}}\\

 &  &  & 4.00 & \mbox{0.7773\,{\scriptsize (0.0058)}} & \mbox{0.5117\,{\scriptsize (0.0058)}} & \mbox{0.0000\,{\scriptsize (0.0000)}} & \mbox{0.5117\,{\scriptsize (0.0058)}} & \textbf{\mbox{0.8304\,{\scriptsize (0.0061)}}} & \mbox{0.6668\,{\scriptsize (0.0063)}}\\

 &  & X & 1.00 & \mbox{0.5222\,{\scriptsize (0.0063)}} & \mbox{0.4058\,{\scriptsize (0.0060)}} & \mbox{0.0000\,{\scriptsize (0.0000)}} & \mbox{0.4057\,{\scriptsize (0.0060)}} & \textbf{\mbox{0.5488\,{\scriptsize (0.0063)}}} & \mbox{0.4640\,{\scriptsize (0.0063)}}\\

 & 0.3 & E & 0.25 & \textbf{\mbox{0.5356\,{\scriptsize (0.0081)}}} & \mbox{0.5042\,{\scriptsize (0.0077)}} & \mbox{0.0000\,{\scriptsize (0.0000)}} & \mbox{0.3957\,{\scriptsize (0.0120)}} & \mbox{0.5206\,{\scriptsize (0.0087)}} & \mbox{0.5330\,{\scriptsize (0.0075)}}\\

 &  &  & 1.00 & \mbox{0.6697\,{\scriptsize (0.0060)}} & \mbox{0.5828\,{\scriptsize (0.0065)}} & \mbox{0.0000\,{\scriptsize (0.0000)}} & \mbox{0.5716\,{\scriptsize (0.0073)}} & \textbf{\mbox{0.7192\,{\scriptsize (0.0065)}}} & \mbox{0.6164\,{\scriptsize (0.0072)}}\\

 &  &  & 4.00 & \mbox{0.8559\,{\scriptsize (0.0052)}} & \mbox{0.5804\,{\scriptsize (0.0065)}} & \mbox{0.0000\,{\scriptsize (0.0000)}} & \mbox{0.5804\,{\scriptsize (0.0065)}} & \textbf{\mbox{0.8671\,{\scriptsize (0.0051)}}} & \mbox{0.7086\,{\scriptsize (0.0056)}}\\

 &  & H & 0.25 & \textbf{\mbox{0.4843\,{\scriptsize (0.0078)}}} & \mbox{0.4505\,{\scriptsize (0.0071)}} & \mbox{0.0000\,{\scriptsize (0.0000)}} & \mbox{0.3831\,{\scriptsize (0.0099)}} & \mbox{0.4701\,{\scriptsize (0.0074)}} & \mbox{0.4704\,{\scriptsize (0.0074)}}\\

 &  &  & 1.00 & \mbox{0.6420\,{\scriptsize (0.0066)}} & \mbox{0.5553\,{\scriptsize (0.0066)}} & \mbox{0.0000\,{\scriptsize (0.0000)}} & \mbox{0.5499\,{\scriptsize (0.0069)}} & \textbf{\mbox{0.6821\,{\scriptsize (0.0064)}}} & \mbox{0.5775\,{\scriptsize (0.0064)}}\\

 &  &  & 4.00 & \mbox{0.7800\,{\scriptsize (0.0061)}} & \mbox{0.5709\,{\scriptsize (0.0062)}} & \mbox{0.0000\,{\scriptsize (0.0000)}} & \mbox{0.5709\,{\scriptsize (0.0062)}} & \textbf{\mbox{0.8280\,{\scriptsize (0.0057)}}} & \mbox{0.6893\,{\scriptsize (0.0065)}}\\

 & 0.6 & E & 0.25 & \mbox{0.4424\,{\scriptsize (0.0072)}} & \textbf{\mbox{0.4845\,{\scriptsize (0.0076)}}} & \mbox{0.0000\,{\scriptsize (0.0000)}} & \mbox{0.4268\,{\scriptsize (0.0099)}} & \mbox{0.4224\,{\scriptsize (0.0063)}} & \mbox{0.4799\,{\scriptsize (0.0078)}}\\

 &  &  & 1.00 & \mbox{0.4987\,{\scriptsize (0.0057)}} & \textbf{\mbox{0.6035\,{\scriptsize (0.0072)}}} & \mbox{0.0000\,{\scriptsize (0.0000)}} & \mbox{0.5656\,{\scriptsize (0.0080)}} & \mbox{0.5163\,{\scriptsize (0.0055)}} & \mbox{0.4995\,{\scriptsize (0.0082)}}\\

 &  &  & 4.00 & \mbox{0.7185\,{\scriptsize (0.0048)}} & \mbox{0.5933\,{\scriptsize (0.0061)}} & \mbox{0.0000\,{\scriptsize (0.0000)}} & \mbox{0.5903\,{\scriptsize (0.0061)}} & \textbf{\mbox{0.7386\,{\scriptsize (0.0048)}}} & \mbox{0.5993\,{\scriptsize (0.0042)}}\\

 &  & H & 0.25 & \mbox{0.3830\,{\scriptsize (0.0076)}} & \textbf{\mbox{0.4443\,{\scriptsize (0.0072)}}} & \mbox{0.0000\,{\scriptsize (0.0000)}} & \mbox{0.4032\,{\scriptsize (0.0084)}} & \mbox{0.3732\,{\scriptsize (0.0064)}} & \mbox{0.4260\,{\scriptsize (0.0086)}}\\

 &  &  & 1.00 & \mbox{0.4914\,{\scriptsize (0.0063)}} & \textbf{\mbox{0.5732\,{\scriptsize (0.0069)}}} & \mbox{0.0000\,{\scriptsize (0.0000)}} & \mbox{0.5547\,{\scriptsize (0.0078)}} & \mbox{0.5059\,{\scriptsize (0.0061)}} & \mbox{0.4808\,{\scriptsize (0.0075)}}\\

 &  &  & 4.00 & \mbox{0.6576\,{\scriptsize (0.0053)}} & \mbox{0.6000\,{\scriptsize (0.0067)}} & \mbox{0.0000\,{\scriptsize (0.0000)}} & \mbox{0.5985\,{\scriptsize (0.0067)}} & \textbf{\mbox{0.6812\,{\scriptsize (0.0049)}}} & \mbox{0.5700\,{\scriptsize (0.0052)}}\\

 & 0.9 & E & 0.25 & \mbox{0.3200\,{\scriptsize (0.0054)}} & \textbf{\mbox{0.3691\,{\scriptsize (0.0066)}}} & \mbox{0.0002\,{\scriptsize (0.0001)}} & \mbox{0.3495\,{\scriptsize (0.0080)}} & \mbox{0.3289\,{\scriptsize (0.0053)}} & \mbox{0.2705\,{\scriptsize (0.0062)}}\\

 &  &  & 1.00 & \mbox{0.4147\,{\scriptsize (0.0048)}} & \textbf{\mbox{0.5021\,{\scriptsize (0.0059)}}} & \mbox{0.0000\,{\scriptsize (0.0000)}} & \mbox{0.4811\,{\scriptsize (0.0070)}} & \mbox{0.4256\,{\scriptsize (0.0046)}} & \mbox{0.3547\,{\scriptsize (0.0046)}}\\

 &  &  & 4.00 & \textbf{\mbox{0.6437\,{\scriptsize (0.0068)}}} & \mbox{0.5611\,{\scriptsize (0.0047)}} & \mbox{0.0000\,{\scriptsize (0.0000)}} & \mbox{0.5465\,{\scriptsize (0.0051)}} & \mbox{0.6235\,{\scriptsize (0.0059)}} & \mbox{0.5261\,{\scriptsize (0.0056)}}\\

 &  & H & 0.25 & \mbox{0.2747\,{\scriptsize (0.0047)}} & \textbf{\mbox{0.3295\,{\scriptsize (0.0073)}}} & \mbox{0.0000\,{\scriptsize (0.0000)}} & \mbox{0.3244\,{\scriptsize (0.0081)}} & \mbox{0.2805\,{\scriptsize (0.0046)}} & \mbox{0.2529\,{\scriptsize (0.0063)}}\\

 &  &  & 1.00 & \mbox{0.3495\,{\scriptsize (0.0048)}} & \textbf{\mbox{0.4146\,{\scriptsize (0.0060)}}} & \mbox{0.0000\,{\scriptsize (0.0000)}} & \mbox{0.4042\,{\scriptsize (0.0069)}} & \mbox{0.3694\,{\scriptsize (0.0053)}} & \mbox{0.2956\,{\scriptsize (0.0056)}}\\

 &  &  & 4.00 & \textbf{\mbox{0.5395\,{\scriptsize (0.0049)}}} & \mbox{0.5266\,{\scriptsize (0.0053)}} & \mbox{0.0000\,{\scriptsize (0.0000)}} & \mbox{0.5123\,{\scriptsize (0.0059)}} & \mbox{0.5308\,{\scriptsize (0.0050)}} & \mbox{0.4501\,{\scriptsize (0.0050)}}\\

 &  & X & 1.00 & \mbox{0.2768\,{\scriptsize (0.0045)}} & \textbf{\mbox{0.4160\,{\scriptsize (0.0078)}}} & \mbox{0.0000\,{\scriptsize (0.0000)}} & \mbox{0.4154\,{\scriptsize (0.0085)}} & \mbox{0.2932\,{\scriptsize (0.0048)}} & \mbox{0.2437\,{\scriptsize (0.0055)}}\\
\cmidrule{1-10}
500 & 0.0 & E & 0.25 & \mbox{0.3460\,{\scriptsize (0.0107)}} & \mbox{0.3266\,{\scriptsize (0.0094)}} & \mbox{0.0001\,{\scriptsize (0.0000)}} & \mbox{0.2599\,{\scriptsize (0.0103)}} & \textbf{\mbox{0.3476\,{\scriptsize (0.0112)}}} & \mbox{0.3308\,{\scriptsize (0.0094)}}\\

 &  &  & 1.00 & \mbox{0.5897\,{\scriptsize (0.0046)}} & \mbox{0.4562\,{\scriptsize (0.0047)}} & \mbox{0.0000\,{\scriptsize (0.0000)}} & \mbox{0.4558\,{\scriptsize (0.0048)}} & \textbf{\mbox{0.6802\,{\scriptsize (0.0055)}}} & \mbox{0.4634\,{\scriptsize (0.0043)}}\\

 &  &  & 4.00 & \mbox{0.8747\,{\scriptsize (0.0063)}} & \mbox{0.4505\,{\scriptsize (0.0049)}} & \mbox{0.0000\,{\scriptsize (0.0000)}} & \mbox{0.4505\,{\scriptsize (0.0049)}} & \textbf{\mbox{0.8849\,{\scriptsize (0.0056)}}} & \mbox{0.6839\,{\scriptsize (0.0063)}}\\

 &  & H & 0.25 & \mbox{0.3417\,{\scriptsize (0.0082)}} & \mbox{0.3202\,{\scriptsize (0.0075)}} & \mbox{0.0001\,{\scriptsize (0.0000)}} & \mbox{0.2819\,{\scriptsize (0.0084)}} & \textbf{\mbox{0.3546\,{\scriptsize (0.0094)}}} & \mbox{0.3224\,{\scriptsize (0.0075)}}\\

 &  &  & 1.00 & \mbox{0.5591\,{\scriptsize (0.0058)}} & \mbox{0.4456\,{\scriptsize (0.0054)}} & \mbox{0.0000\,{\scriptsize (0.0000)}} & \mbox{0.4456\,{\scriptsize (0.0054)}} & \textbf{\mbox{0.6226\,{\scriptsize (0.0062)}}} & \mbox{0.4587\,{\scriptsize (0.0049)}}\\

 &  &  & 4.00 & \mbox{0.7124\,{\scriptsize (0.0066)}} & \mbox{0.4604\,{\scriptsize (0.0051)}} & \mbox{0.0000\,{\scriptsize (0.0000)}} & \mbox{0.4604\,{\scriptsize (0.0051)}} & \textbf{\mbox{0.7966\,{\scriptsize (0.0064)}}} & \mbox{0.5790\,{\scriptsize (0.0064)}}\\

 &  & X & 1.00 & \mbox{0.4562\,{\scriptsize (0.0059)}} & \mbox{0.3476\,{\scriptsize (0.0053)}} & \mbox{0.0000\,{\scriptsize (0.0000)}} & \mbox{0.3476\,{\scriptsize (0.0053)}} & \textbf{\mbox{0.4931\,{\scriptsize (0.0058)}}} & \mbox{0.3741\,{\scriptsize (0.0053)}}\\

 & 0.3 & E & 0.25 & \textbf{\mbox{0.4368\,{\scriptsize (0.0082)}}} & \mbox{0.4182\,{\scriptsize (0.0079)}} & \mbox{0.0000\,{\scriptsize (0.0000)}} & \mbox{0.3638\,{\scriptsize (0.0095)}} & \mbox{0.4258\,{\scriptsize (0.0084)}} & \mbox{0.4222\,{\scriptsize (0.0077)}}\\

 &  &  & 1.00 & \mbox{0.6207\,{\scriptsize (0.0076)}} & \mbox{0.5205\,{\scriptsize (0.0062)}} & \mbox{0.0000\,{\scriptsize (0.0000)}} & \mbox{0.5145\,{\scriptsize (0.0065)}} & \textbf{\mbox{0.6870\,{\scriptsize (0.0068)}}} & \mbox{0.5780\,{\scriptsize (0.0078)}}\\

 &  &  & 4.00 & \mbox{0.7188\,{\scriptsize (0.0057)}} & \mbox{0.5085\,{\scriptsize (0.0057)}} & \mbox{0.0000\,{\scriptsize (0.0000)}} & \mbox{0.5085\,{\scriptsize (0.0057)}} & \textbf{\mbox{0.7633\,{\scriptsize (0.0057)}}} & \mbox{0.5358\,{\scriptsize (0.0046)}}\\

 &  & H & 0.25 & \mbox{0.4056\,{\scriptsize (0.0073)}} & \mbox{0.3792\,{\scriptsize (0.0068)}} & \mbox{0.0000\,{\scriptsize (0.0000)}} & \mbox{0.3392\,{\scriptsize (0.0087)}} & \textbf{\mbox{0.4084\,{\scriptsize (0.0072)}}} & \mbox{0.3804\,{\scriptsize (0.0069)}}\\

 &  &  & 1.00 & \mbox{0.5691\,{\scriptsize (0.0064)}} & \mbox{0.4979\,{\scriptsize (0.0065)}} & \mbox{0.0000\,{\scriptsize (0.0000)}} & \mbox{0.4961\,{\scriptsize (0.0066)}} & \textbf{\mbox{0.6187\,{\scriptsize (0.0065)}}} & \mbox{0.5168\,{\scriptsize (0.0073)}}\\

 &  &  & 4.00 & \mbox{0.7245\,{\scriptsize (0.0071)}} & \mbox{0.5272\,{\scriptsize (0.0062)}} & \mbox{0.0000\,{\scriptsize (0.0000)}} & \mbox{0.5272\,{\scriptsize (0.0062)}} & \textbf{\mbox{0.7877\,{\scriptsize (0.0064)}}} & \mbox{0.5942\,{\scriptsize (0.0067)}}\\

 & 0.6 & E & 0.25 & \mbox{0.3865\,{\scriptsize (0.0076)}} & \textbf{\mbox{0.4168\,{\scriptsize (0.0074)}}} & \mbox{0.0000\,{\scriptsize (0.0000)}} & \mbox{0.3860\,{\scriptsize (0.0083)}} & \mbox{0.3789\,{\scriptsize (0.0064)}} & \mbox{0.4058\,{\scriptsize (0.0078)}}\\

 &  &  & 1.00 & \mbox{0.4206\,{\scriptsize (0.0061)}} & \textbf{\mbox{0.5342\,{\scriptsize (0.0071)}}} & \mbox{0.0000\,{\scriptsize (0.0000)}} & \mbox{0.5175\,{\scriptsize (0.0074)}} & \mbox{0.4753\,{\scriptsize (0.0063)}} & \mbox{0.5029\,{\scriptsize (0.0117)}}\\

 &  &  & 4.00 & \mbox{0.5665\,{\scriptsize (0.0042)}} & \mbox{0.5554\,{\scriptsize (0.0063)}} & \mbox{0.0000\,{\scriptsize (0.0000)}} & \mbox{0.5545\,{\scriptsize (0.0064)}} & \textbf{\mbox{0.6052\,{\scriptsize (0.0047)}}} & \mbox{0.4515\,{\scriptsize (0.0039)}}\\

 &  & H & 0.25 & \mbox{0.3488\,{\scriptsize (0.0071)}} & \textbf{\mbox{0.3924\,{\scriptsize (0.0070)}}} & \mbox{0.0001\,{\scriptsize (0.0001)}} & \mbox{0.3703\,{\scriptsize (0.0079)}} & \mbox{0.3392\,{\scriptsize (0.0061)}} & \mbox{0.3743\,{\scriptsize (0.0078)}}\\

 &  &  & 1.00 & \mbox{0.4239\,{\scriptsize (0.0064)}} & \textbf{\mbox{0.5061\,{\scriptsize (0.0069)}}} & \mbox{0.0000\,{\scriptsize (0.0000)}} & \mbox{0.4951\,{\scriptsize (0.0071)}} & \mbox{0.4536\,{\scriptsize (0.0058)}} & \mbox{0.4270\,{\scriptsize (0.0089)}}\\

 &  &  & 4.00 & \mbox{0.5394\,{\scriptsize (0.0055)}} & \mbox{0.5430\,{\scriptsize (0.0062)}} & \mbox{0.0000\,{\scriptsize (0.0000)}} & \mbox{0.5422\,{\scriptsize (0.0062)}} & \textbf{\mbox{0.5775\,{\scriptsize (0.0053)}}} & \mbox{0.4592\,{\scriptsize (0.0060)}}\\

 & 0.9 & E & 0.25 & \mbox{0.2592\,{\scriptsize (0.0057)}} & \mbox{0.2859\,{\scriptsize (0.0062)}} & \mbox{0.0005\,{\scriptsize (0.0001)}} & \mbox{0.2854\,{\scriptsize (0.0061)}} & \textbf{\mbox{0.2882\,{\scriptsize (0.0058)}}} & \mbox{0.2107\,{\scriptsize (0.0061)}}\\

 &  &  & 1.00 & \mbox{0.3476\,{\scriptsize (0.0052)}} & \textbf{\mbox{0.4209\,{\scriptsize (0.0052)}}} & \mbox{0.0000\,{\scriptsize (0.0000)}} & \mbox{0.4153\,{\scriptsize (0.0053)}} & \mbox{0.3712\,{\scriptsize (0.0048)}} & \mbox{0.3084\,{\scriptsize (0.0060)}}\\

 &  &  & 4.00 & \mbox{0.4804\,{\scriptsize (0.0061)}} & \mbox{0.4692\,{\scriptsize (0.0044)}} & \mbox{0.0000\,{\scriptsize (0.0000)}} & \mbox{0.4667\,{\scriptsize (0.0045)}} & \textbf{\mbox{0.5050\,{\scriptsize (0.0057)}}} & \mbox{0.3706\,{\scriptsize (0.0050)}}\\

 &  & H & 0.25 & \mbox{0.2554\,{\scriptsize (0.0046)}} & \textbf{\mbox{0.3069\,{\scriptsize (0.0072)}}} & \mbox{-0.0008\,{\scriptsize (0.0010)}} & \mbox{0.3025\,{\scriptsize (0.0071)}} & \mbox{0.2689\,{\scriptsize (0.0042)}} & \mbox{0.2338\,{\scriptsize (0.0071)}}\\

 &  &  & 1.00 & \mbox{0.2981\,{\scriptsize (0.0050)}} & \textbf{\mbox{0.3721\,{\scriptsize (0.0058)}}} & \mbox{0.0000\,{\scriptsize (0.0000)}} & \mbox{0.3676\,{\scriptsize (0.0060)}} & \mbox{0.3149\,{\scriptsize (0.0050)}} & \mbox{0.2545\,{\scriptsize (0.0059)}}\\

 &  &  & 4.00 & \mbox{0.4293\,{\scriptsize (0.0052)}} & \textbf{\mbox{0.4471\,{\scriptsize (0.0050)}}} & \mbox{0.0000\,{\scriptsize (0.0000)}} & \mbox{0.4430\,{\scriptsize (0.0050)}} & \mbox{0.4431\,{\scriptsize (0.0051)}} & \mbox{0.3355\,{\scriptsize (0.0056)}}\\

 &  & X & 1.00 & \mbox{0.2595\,{\scriptsize (0.0044)}} & \textbf{\mbox{0.4107\,{\scriptsize (0.0079)}}} & \mbox{0.0000\,{\scriptsize (0.0000)}} & \mbox{0.4063\,{\scriptsize (0.0079)}} & \mbox{0.2700\,{\scriptsize (0.0042)}} & \mbox{0.2389\,{\scriptsize (0.0055)}}\\
\bottomrule
\end{tabular}
}
\end{table}

Table~\ref{tab:mcc_combined} gives Lambert a clearer selection advantage
over SCAD than over MCP: MCC is higher in 41 main scenarios against SCAD,
but only ten against MCP. For equal signals with SNR 4 and $\rho=0$,
Lambert's MCC is 0.9233 at $p=125$ and 0.8747 at $p=500$, versus SCAD's
0.8153 and 0.6839. MCP slightly improves on Lambert in both settings.
At $p=500$ and $\rho=0.9$, Lambert reaches 0.4804, exceeding SCAD's
0.3706 but falling below MCP's 0.5050.

Lambert also improves on LASSO and ENET in 30 and 32 main scenarios,
respectively, but this advantage depends on the signal structure. Under
extreme heterogeneity at $p=500$ and $\rho=0.9$, LASSO and ENET attain
0.4107 and 0.4063, compared with 0.2595 for Lambert, 0.2700 for MCP,
and 0.2389 for SCAD. Across all stress scenarios, Lambert improves on
SCAD in both NMSE and MCC, while MCP retains higher MCC despite its larger
NMSE. Ridge's near-zero MCC reflects its dense fits, not its usefulness
for prediction.

\begin{table}[!tbp]
\captionsetup{aboveskip=0pt,belowskip=3pt}
\centering\spacingset{1}\scriptsize
\setlength{\tabcolsep}{2pt}
\renewcommand{\arraystretch}{0.90}
\caption{Accuracy of regression-coefficient estimation across signal profiles and predictor correlations: squared coefficient error (CE).}
\label{tab:supp_coefficient_error}
\resizebox{\textwidth}{!}{
\begin{tabular}{>{\centering\arraybackslash}m{0.55cm}>{\centering\arraybackslash}m{0.5cm}>{\centering\arraybackslash}m{0.9cm}>{\centering\arraybackslash}m{0.7cm}>{\raggedleft\arraybackslash}m{2.28cm}>{\raggedleft\arraybackslash}m{2.28cm}>{\raggedleft\arraybackslash}m{2.28cm}>{\raggedleft\arraybackslash}m{2.28cm}>{\raggedleft\arraybackslash}m{2.28cm}>{\raggedleft\arraybackslash}m{2.28cm}}
\toprule
$p$ & $\rho$ & Signal & SNR & Lambert & LASSO & Ridge & ENET & MCP & SCAD\\
\midrule
125 & 0.0 & E & 0.25 & \mbox{0.7107\,{\scriptsize (0.0099)}} & \mbox{0.6421\,{\scriptsize (0.0103)}} & \mbox{0.7465\,{\scriptsize (0.0051)}} & \textbf{\mbox{0.6411\,{\scriptsize (0.0089)}}} & \mbox{0.7093\,{\scriptsize (0.0103)}} & \mbox{0.6639\,{\scriptsize (0.0098)}}\\

 &  &  & 1.00 & \textbf{\mbox{0.1097\,{\scriptsize (0.0034)}}} & \mbox{0.1758\,{\scriptsize (0.0033)}} & \mbox{0.4320\,{\scriptsize (0.0033)}} & \mbox{0.1770\,{\scriptsize (0.0034)}} & \mbox{0.1186\,{\scriptsize (0.0032)}} & \mbox{0.1255\,{\scriptsize (0.0033)}}\\

 &  &  & 4.00 & \textbf{\mbox{0.0115\,{\scriptsize (0.0003)}}} & \mbox{0.0427\,{\scriptsize (0.0008)}} & \mbox{0.1805\,{\scriptsize (0.0016)}} & \mbox{0.0427\,{\scriptsize (0.0008)}} & \mbox{0.0121\,{\scriptsize (0.0004)}} & \mbox{0.0122\,{\scriptsize (0.0003)}}\\

 &  & H & 0.25 & \mbox{0.5943\,{\scriptsize (0.0104)}} & \textbf{\mbox{0.5604\,{\scriptsize (0.0092)}}} & \mbox{0.7520\,{\scriptsize (0.0051)}} & \mbox{0.5775\,{\scriptsize (0.0091)}} & \mbox{0.5908\,{\scriptsize (0.0108)}} & \mbox{0.5661\,{\scriptsize (0.0094)}}\\

 &  &  & 1.00 & \textbf{\mbox{0.1288\,{\scriptsize (0.0031)}}} & \mbox{0.1658\,{\scriptsize (0.0030)}} & \mbox{0.4279\,{\scriptsize (0.0032)}} & \mbox{0.1662\,{\scriptsize (0.0030)}} & \mbox{0.1317\,{\scriptsize (0.0031)}} & \mbox{0.1317\,{\scriptsize (0.0029)}}\\

 &  &  & 4.00 & \textbf{\mbox{0.0196\,{\scriptsize (0.0006)}}} & \mbox{0.0424\,{\scriptsize (0.0008)}} & \mbox{0.1807\,{\scriptsize (0.0015)}} & \mbox{0.0424\,{\scriptsize (0.0008)}} & \mbox{0.0204\,{\scriptsize (0.0006)}} & \mbox{0.0223\,{\scriptsize (0.0006)}}\\

 &  & X & 1.00 & \textbf{\mbox{0.0900\,{\scriptsize (0.0021)}}} & \mbox{0.1206\,{\scriptsize (0.0023)}} & \mbox{0.4267\,{\scriptsize (0.0030)}} & \mbox{0.1206\,{\scriptsize (0.0023)}} & \mbox{0.0908\,{\scriptsize (0.0021)}} & \mbox{0.0907\,{\scriptsize (0.0019)}}\\

 & 0.3 & E & 0.25 & \mbox{0.7691\,{\scriptsize (0.0130)}} & \mbox{0.6259\,{\scriptsize (0.0082)}} & \mbox{0.6990\,{\scriptsize (0.0056)}} & \textbf{\mbox{0.6221\,{\scriptsize (0.0084)}}} & \mbox{0.7810\,{\scriptsize (0.0120)}} & \mbox{0.6844\,{\scriptsize (0.0114)}}\\

 &  &  & 1.00 & \mbox{0.3222\,{\scriptsize (0.0079)}} & \textbf{\mbox{0.2095\,{\scriptsize (0.0040)}}} & \mbox{0.4230\,{\scriptsize (0.0033)}} & \mbox{0.2124\,{\scriptsize (0.0042)}} & \mbox{0.3112\,{\scriptsize (0.0073)}} & \mbox{0.3185\,{\scriptsize (0.0064)}}\\

 &  &  & 4.00 & \textbf{\mbox{0.0220\,{\scriptsize (0.0007)}}} & \mbox{0.0532\,{\scriptsize (0.0012)}} & \mbox{0.2128\,{\scriptsize (0.0019)}} & \mbox{0.0532\,{\scriptsize (0.0012)}} & \mbox{0.0245\,{\scriptsize (0.0008)}} & \mbox{0.0253\,{\scriptsize (0.0008)}}\\

 &  & H & 0.25 & \mbox{0.6731\,{\scriptsize (0.0112)}} & \textbf{\mbox{0.5514\,{\scriptsize (0.0087)}}} & \mbox{0.7089\,{\scriptsize (0.0053)}} & \mbox{0.5603\,{\scriptsize (0.0085)}} & \mbox{0.6676\,{\scriptsize (0.0107)}} & \mbox{0.5995\,{\scriptsize (0.0097)}}\\

 &  &  & 1.00 & \mbox{0.2459\,{\scriptsize (0.0059)}} & \textbf{\mbox{0.1871\,{\scriptsize (0.0033)}}} & \mbox{0.4278\,{\scriptsize (0.0029)}} & \mbox{0.1886\,{\scriptsize (0.0033)}} & \mbox{0.2374\,{\scriptsize (0.0057)}} & \mbox{0.2353\,{\scriptsize (0.0052)}}\\

 &  &  & 4.00 & \mbox{0.0482\,{\scriptsize (0.0013)}} & \mbox{0.0529\,{\scriptsize (0.0011)}} & \mbox{0.2112\,{\scriptsize (0.0017)}} & \mbox{0.0529\,{\scriptsize (0.0011)}} & \textbf{\mbox{0.0455\,{\scriptsize (0.0013)}}} & \mbox{0.0465\,{\scriptsize (0.0011)}}\\

 & 0.6 & E & 0.25 & \mbox{1.1560\,{\scriptsize (0.0235)}} & \mbox{0.7177\,{\scriptsize (0.0107)}} & \mbox{0.6661\,{\scriptsize (0.0060)}} & \textbf{\mbox{0.6603\,{\scriptsize (0.0096)}}} & \mbox{1.1554\,{\scriptsize (0.0222)}} & \mbox{0.9151\,{\scriptsize (0.0211)}}\\

 &  &  & 1.00 & \mbox{0.8990\,{\scriptsize (0.0120)}} & \mbox{0.3492\,{\scriptsize (0.0061)}} & \mbox{0.4384\,{\scriptsize (0.0044)}} & \textbf{\mbox{0.3255\,{\scriptsize (0.0059)}}} & \mbox{0.9011\,{\scriptsize (0.0115)}} & \mbox{0.8447\,{\scriptsize (0.0111)}}\\

 &  &  & 4.00 & \mbox{0.1097\,{\scriptsize (0.0065)}} & \mbox{0.1021\,{\scriptsize (0.0025)}} & \mbox{0.2450\,{\scriptsize (0.0022)}} & \textbf{\mbox{0.1015\,{\scriptsize (0.0025)}}} & \mbox{0.1135\,{\scriptsize (0.0066)}} & \mbox{0.1088\,{\scriptsize (0.0061)}}\\

 &  & H & 0.25 & \mbox{1.1056\,{\scriptsize (0.0250)}} & \mbox{0.6421\,{\scriptsize (0.0117)}} & \mbox{0.6727\,{\scriptsize (0.0065)}} & \textbf{\mbox{0.6119\,{\scriptsize (0.0106)}}} & \mbox{1.0738\,{\scriptsize (0.0223)}} & \mbox{0.9018\,{\scriptsize (0.0248)}}\\

 &  &  & 1.00 & \mbox{0.7205\,{\scriptsize (0.0132)}} & \mbox{0.3154\,{\scriptsize (0.0050)}} & \mbox{0.4368\,{\scriptsize (0.0041)}} & \textbf{\mbox{0.3050\,{\scriptsize (0.0051)}}} & \mbox{0.7160\,{\scriptsize (0.0132)}} & \mbox{0.6842\,{\scriptsize (0.0117)}}\\

 &  &  & 4.00 & \mbox{0.1541\,{\scriptsize (0.0052)}} & \mbox{0.1055\,{\scriptsize (0.0027)}} & \mbox{0.2465\,{\scriptsize (0.0022)}} & \textbf{\mbox{0.1053\,{\scriptsize (0.0027)}}} & \mbox{0.1586\,{\scriptsize (0.0053)}} & \mbox{0.1515\,{\scriptsize (0.0050)}}\\

 & 0.9 & E & 0.25 & \mbox{1.9702\,{\scriptsize (0.0486)}} & \mbox{0.9126\,{\scriptsize (0.0211)}} & \textbf{\mbox{0.7124\,{\scriptsize (0.0146)}}} & \mbox{0.8133\,{\scriptsize (0.0192)}} & \mbox{1.8812\,{\scriptsize (0.0512)}} & \mbox{1.9087\,{\scriptsize (0.0458)}}\\

 &  &  & 1.00 & \mbox{1.4411\,{\scriptsize (0.0199)}} & \mbox{0.5521\,{\scriptsize (0.0082)}} & \mbox{0.4804\,{\scriptsize (0.0069)}} & \textbf{\mbox{0.4760\,{\scriptsize (0.0098)}}} & \mbox{1.4907\,{\scriptsize (0.0190)}} & \mbox{1.5124\,{\scriptsize (0.0200)}}\\

 &  &  & 4.00 & \mbox{0.5484\,{\scriptsize (0.0181)}} & \mbox{0.2451\,{\scriptsize (0.0047)}} & \mbox{0.2956\,{\scriptsize (0.0031)}} & \textbf{\mbox{0.2239\,{\scriptsize (0.0045)}}} & \mbox{0.6340\,{\scriptsize (0.0171)}} & \mbox{0.6226\,{\scriptsize (0.0169)}}\\

 &  & H & 0.25 & \mbox{1.6524\,{\scriptsize (0.0616)}} & \mbox{0.8561\,{\scriptsize (0.0223)}} & \textbf{\mbox{0.7229\,{\scriptsize (0.0155)}}} & \mbox{0.7715\,{\scriptsize (0.0206)}} & \mbox{1.6153\,{\scriptsize (0.0653)}} & \mbox{1.5874\,{\scriptsize (0.0597)}}\\

 &  &  & 1.00 & \mbox{1.4535\,{\scriptsize (0.0199)}} & \mbox{0.5609\,{\scriptsize (0.0099)}} & \textbf{\mbox{0.5105\,{\scriptsize (0.0092)}}} & \mbox{0.5110\,{\scriptsize (0.0118)}} & \mbox{1.3996\,{\scriptsize (0.0203)}} & \mbox{1.4121\,{\scriptsize (0.0200)}}\\

 &  &  & 4.00 & \mbox{0.6139\,{\scriptsize (0.0153)}} & \mbox{0.2909\,{\scriptsize (0.0043)}} & \mbox{0.3200\,{\scriptsize (0.0035)}} & \textbf{\mbox{0.2696\,{\scriptsize (0.0044)}}} & \mbox{0.6581\,{\scriptsize (0.0155)}} & \mbox{0.6572\,{\scriptsize (0.0161)}}\\

 &  & X & 1.00 & \mbox{1.3907\,{\scriptsize (0.0335)}} & \mbox{0.3431\,{\scriptsize (0.0119)}} & \mbox{0.4740\,{\scriptsize (0.0087)}} & \textbf{\mbox{0.2872\,{\scriptsize (0.0112)}}} & \mbox{1.2539\,{\scriptsize (0.0311)}} & \mbox{1.3182\,{\scriptsize (0.0317)}}\\
\cmidrule{1-10}
500 & 0.0 & E & 0.25 & \mbox{0.8349\,{\scriptsize (0.0080)}} & \mbox{0.8244\,{\scriptsize (0.0088)}} & \mbox{0.9470\,{\scriptsize (0.0031)}} & \mbox{0.8362\,{\scriptsize (0.0081)}} & \mbox{0.8470\,{\scriptsize (0.0081)}} & \textbf{\mbox{0.8200\,{\scriptsize (0.0082)}}}\\

 &  &  & 1.00 & \textbf{\mbox{0.1764\,{\scriptsize (0.0050)}}} & \mbox{0.2717\,{\scriptsize (0.0046)}} & \mbox{0.8023\,{\scriptsize (0.0025)}} & \mbox{0.2726\,{\scriptsize (0.0047)}} & \mbox{0.1805\,{\scriptsize (0.0048)}} & \mbox{0.2084\,{\scriptsize (0.0049)}}\\

 &  &  & 4.00 & \textbf{\mbox{0.0121\,{\scriptsize (0.0003)}}} & \mbox{0.0688\,{\scriptsize (0.0012)}} & \mbox{0.6569\,{\scriptsize (0.0021)}} & \mbox{0.0688\,{\scriptsize (0.0012)}} & \mbox{0.0133\,{\scriptsize (0.0004)}} & \mbox{0.0146\,{\scriptsize (0.0004)}}\\

 &  & H & 0.25 & \textbf{\mbox{0.7034\,{\scriptsize (0.0103)}}} & \mbox{0.7125\,{\scriptsize (0.0096)}} & \mbox{0.9527\,{\scriptsize (0.0039)}} & \mbox{0.7306\,{\scriptsize (0.0100)}} & \mbox{0.7102\,{\scriptsize (0.0105)}} & \mbox{0.7055\,{\scriptsize (0.0096)}}\\

 &  &  & 1.00 & \textbf{\mbox{0.1857\,{\scriptsize (0.0038)}}} & \mbox{0.2565\,{\scriptsize (0.0039)}} & \mbox{0.8012\,{\scriptsize (0.0023)}} & \mbox{0.2565\,{\scriptsize (0.0039)}} & \mbox{0.1874\,{\scriptsize (0.0038)}} & \mbox{0.2004\,{\scriptsize (0.0038)}}\\

 &  &  & 4.00 & \mbox{0.0296\,{\scriptsize (0.0008)}} & \mbox{0.0697\,{\scriptsize (0.0011)}} & \mbox{0.6545\,{\scriptsize (0.0020)}} & \mbox{0.0697\,{\scriptsize (0.0011)}} & \textbf{\mbox{0.0290\,{\scriptsize (0.0008)}}} & \mbox{0.0337\,{\scriptsize (0.0008)}}\\

 &  & X & 1.00 & \textbf{\mbox{0.1187\,{\scriptsize (0.0027)}}} & \mbox{0.1768\,{\scriptsize (0.0030)}} & \mbox{0.8017\,{\scriptsize (0.0023)}} & \mbox{0.1768\,{\scriptsize (0.0030)}} & \mbox{0.1197\,{\scriptsize (0.0029)}} & \mbox{0.1263\,{\scriptsize (0.0026)}}\\

 & 0.3 & E & 0.25 & \mbox{0.8179\,{\scriptsize (0.0091)}} & \textbf{\mbox{0.7619\,{\scriptsize (0.0076)}}} & \mbox{0.9373\,{\scriptsize (0.0060)}} & \mbox{0.7732\,{\scriptsize (0.0095)}} & \mbox{0.8405\,{\scriptsize (0.0085)}} & \mbox{0.7690\,{\scriptsize (0.0080)}}\\

 &  &  & 1.00 & \mbox{0.4686\,{\scriptsize (0.0069)}} & \textbf{\mbox{0.2982\,{\scriptsize (0.0045)}}} & \mbox{0.7590\,{\scriptsize (0.0032)}} & \mbox{0.3001\,{\scriptsize (0.0046)}} & \mbox{0.4220\,{\scriptsize (0.0073)}} & \mbox{0.4173\,{\scriptsize (0.0063)}}\\

 &  &  & 4.00 & \textbf{\mbox{0.0276\,{\scriptsize (0.0009)}}} & \mbox{0.0764\,{\scriptsize (0.0015)}} & \mbox{0.6022\,{\scriptsize (0.0028)}} & \mbox{0.0764\,{\scriptsize (0.0015)}} & \mbox{0.0329\,{\scriptsize (0.0011)}} & \mbox{0.0353\,{\scriptsize (0.0010)}}\\

 &  & H & 0.25 & \mbox{0.7083\,{\scriptsize (0.0094)}} & \textbf{\mbox{0.6635\,{\scriptsize (0.0097)}}} & \mbox{0.9295\,{\scriptsize (0.0050)}} & \mbox{0.6798\,{\scriptsize (0.0105)}} & \mbox{0.7151\,{\scriptsize (0.0089)}} & \mbox{0.6764\,{\scriptsize (0.0097)}}\\

 &  &  & 1.00 & \mbox{0.3483\,{\scriptsize (0.0069)}} & \textbf{\mbox{0.2753\,{\scriptsize (0.0043)}}} & \mbox{0.7695\,{\scriptsize (0.0036)}} & \mbox{0.2763\,{\scriptsize (0.0044)}} & \mbox{0.3338\,{\scriptsize (0.0068)}} & \mbox{0.3358\,{\scriptsize (0.0059)}}\\

 &  &  & 4.00 & \mbox{0.0579\,{\scriptsize (0.0013)}} & \mbox{0.0740\,{\scriptsize (0.0012)}} & \mbox{0.6030\,{\scriptsize (0.0029)}} & \mbox{0.0740\,{\scriptsize (0.0012)}} & \textbf{\mbox{0.0539\,{\scriptsize (0.0014)}}} & \mbox{0.0566\,{\scriptsize (0.0013)}}\\

 & 0.6 & E & 0.25 & \mbox{1.0655\,{\scriptsize (0.0211)}} & \mbox{0.7856\,{\scriptsize (0.0108)}} & \mbox{0.8971\,{\scriptsize (0.0049)}} & \textbf{\mbox{0.7662\,{\scriptsize (0.0101)}}} & \mbox{1.0748\,{\scriptsize (0.0182)}} & \mbox{0.8801\,{\scriptsize (0.0184)}}\\

 &  &  & 1.00 & \mbox{1.0198\,{\scriptsize (0.0106)}} & \mbox{0.4035\,{\scriptsize (0.0056)}} & \mbox{0.7232\,{\scriptsize (0.0045)}} & \textbf{\mbox{0.3952\,{\scriptsize (0.0055)}}} & \mbox{1.0001\,{\scriptsize (0.0111)}} & \mbox{0.8771\,{\scriptsize (0.0152)}}\\

 &  &  & 4.00 & \mbox{0.2418\,{\scriptsize (0.0121)}} & \mbox{0.1521\,{\scriptsize (0.0032)}} & \mbox{0.5713\,{\scriptsize (0.0041)}} & \textbf{\mbox{0.1521\,{\scriptsize (0.0032)}}} & \mbox{0.2810\,{\scriptsize (0.0120)}} & \mbox{0.2641\,{\scriptsize (0.0114)}}\\

 &  & H & 0.25 & \mbox{0.9892\,{\scriptsize (0.0193)}} & \mbox{0.6773\,{\scriptsize (0.0099)}} & \mbox{0.9000\,{\scriptsize (0.0068)}} & \textbf{\mbox{0.6727\,{\scriptsize (0.0106)}}} & \mbox{0.9894\,{\scriptsize (0.0177)}} & \mbox{0.7974\,{\scriptsize (0.0181)}}\\

 &  &  & 1.00 & \mbox{0.8469\,{\scriptsize (0.0115)}} & \mbox{0.3723\,{\scriptsize (0.0045)}} & \mbox{0.7175\,{\scriptsize (0.0041)}} & \textbf{\mbox{0.3696\,{\scriptsize (0.0045)}}} & \mbox{0.8303\,{\scriptsize (0.0116)}} & \mbox{0.7850\,{\scriptsize (0.0114)}}\\

 &  &  & 4.00 & \mbox{0.2514\,{\scriptsize (0.0078)}} & \mbox{0.1525\,{\scriptsize (0.0032)}} & \mbox{0.5776\,{\scriptsize (0.0038)}} & \textbf{\mbox{0.1524\,{\scriptsize (0.0031)}}} & \mbox{0.2760\,{\scriptsize (0.0079)}} & \mbox{0.2689\,{\scriptsize (0.0077)}}\\

 & 0.9 & E & 0.25 & \mbox{1.8355\,{\scriptsize (0.0350)}} & \mbox{0.8836\,{\scriptsize (0.0102)}} & \mbox{0.9307\,{\scriptsize (0.0052)}} & \textbf{\mbox{0.8527\,{\scriptsize (0.0094)}}} & \mbox{1.6824\,{\scriptsize (0.0381)}} & \mbox{1.6942\,{\scriptsize (0.0400)}}\\

 &  &  & 1.00 & \mbox{1.6397\,{\scriptsize (0.0213)}} & \mbox{0.5957\,{\scriptsize (0.0078)}} & \mbox{0.6973\,{\scriptsize (0.0053)}} & \textbf{\mbox{0.5671\,{\scriptsize (0.0086)}}} & \mbox{1.6767\,{\scriptsize (0.0208)}} & \mbox{1.7191\,{\scriptsize (0.0205)}}\\

 &  &  & 4.00 & \mbox{0.8455\,{\scriptsize (0.0173)}} & \mbox{0.3200\,{\scriptsize (0.0040)}} & \mbox{0.5825\,{\scriptsize (0.0052)}} & \textbf{\mbox{0.3114\,{\scriptsize (0.0039)}}} & \mbox{0.9492\,{\scriptsize (0.0155)}} & \mbox{0.9232\,{\scriptsize (0.0156)}}\\

 &  & H & 0.25 & \mbox{1.3610\,{\scriptsize (0.0296)}} & \mbox{0.7698\,{\scriptsize (0.0124)}} & \mbox{0.8740\,{\scriptsize (0.0059)}} & \textbf{\mbox{0.7288\,{\scriptsize (0.0116)}}} & \mbox{1.3092\,{\scriptsize (0.0341)}} & \mbox{1.2701\,{\scriptsize (0.0348)}}\\

 &  &  & 1.00 & \mbox{1.4679\,{\scriptsize (0.0201)}} & \mbox{0.5488\,{\scriptsize (0.0084)}} & \mbox{0.6908\,{\scriptsize (0.0058)}} & \textbf{\mbox{0.5208\,{\scriptsize (0.0086)}}} & \mbox{1.4235\,{\scriptsize (0.0202)}} & \mbox{1.4324\,{\scriptsize (0.0201)}}\\

 &  &  & 4.00 & \mbox{0.8583\,{\scriptsize (0.0189)}} & \mbox{0.3432\,{\scriptsize (0.0040)}} & \mbox{0.5805\,{\scriptsize (0.0050)}} & \textbf{\mbox{0.3308\,{\scriptsize (0.0039)}}} & \mbox{0.9416\,{\scriptsize (0.0174)}} & \mbox{0.9059\,{\scriptsize (0.0183)}}\\

 &  & X & 1.00 & \mbox{1.6027\,{\scriptsize (0.0299)}} & \mbox{0.3387\,{\scriptsize (0.0105)}} & \mbox{0.6633\,{\scriptsize (0.0056)}} & \textbf{\mbox{0.3064\,{\scriptsize (0.0105)}}} & \mbox{1.4214\,{\scriptsize (0.0309)}} & \mbox{1.4872\,{\scriptsize (0.0303)}}\\
\bottomrule
\end{tabular}
}
\end{table}

Table~\ref{tab:supp_coefficient_error} favors Lambert for strong,
independent equal signals. At SNR 4, its CE is 0.0115 for $p=125$ and
0.0121 for $p=500$, compared with 0.0122 and 0.0146 for SCAD. MCP gives
0.0121 and 0.0133, whereas LASSO gives 0.0427 and 0.0688. This advantage
is not general: Lambert has smaller CE than SCAD in only 19 main scenarios,
despite its more frequent gains in NMSE and MCC.

Strong correlation particularly favors quadratic shrinkage. For equal
signals at $p=500$, $\rho=0.9$, and SNR 4, ENET has CE 0.3114, compared
with 0.8455 for Lambert, 0.9492 for MCP, and 0.9232 for SCAD. At
$p=125$, $\rho=0.9$, and SNR 0.25, ridge achieves 0.7124 versus Lambert's
1.9702. These differences are substantial because the true coefficient
vector has squared norm one. Correlated predictors allow errors in
individual coefficients to offset within the linear predictor. Consequently,
a method can retain useful predictions while estimating coefficients poorly;
Lambert's prediction and selection advantages do not imply uniformly
better coefficient estimation.

\begin{table}[!tbp]
\captionsetup{aboveskip=0pt,belowskip=3pt}
\centering\spacingset{1}\scriptsize
\setlength{\tabcolsep}{2pt}
\renewcommand{\arraystretch}{0.90}
\caption{Prediction loss due to estimation across signal profiles and predictor correlations: normalized excess risk (ER).}
\label{tab:supp_excess_risk}
\resizebox{\textwidth}{!}{
\begin{tabular}{>{\centering\arraybackslash}m{0.55cm}>{\centering\arraybackslash}m{0.5cm}>{\centering\arraybackslash}m{0.9cm}>{\centering\arraybackslash}m{0.7cm}>{\raggedleft\arraybackslash}m{2.28cm}>{\raggedleft\arraybackslash}m{2.28cm}>{\raggedleft\arraybackslash}m{2.28cm}>{\raggedleft\arraybackslash}m{2.28cm}>{\raggedleft\arraybackslash}m{2.28cm}>{\raggedleft\arraybackslash}m{2.28cm}}
\toprule
$p$ & $\rho$ & Signal & SNR & Lambert & LASSO & Ridge & ENET & MCP & SCAD\\
\midrule
125 & 0.0 & E & 0.25 & \mbox{0.1827\,{\scriptsize (0.0025)}} & \mbox{0.1653\,{\scriptsize (0.0026)}} & \mbox{0.1915\,{\scriptsize (0.0013)}} & \textbf{\mbox{0.1651\,{\scriptsize (0.0023)}}} & \mbox{0.1823\,{\scriptsize (0.0026)}} & \mbox{0.1709\,{\scriptsize (0.0025)}}\\

 &  &  & 1.00 & \textbf{\mbox{0.1145\,{\scriptsize (0.0034)}}} & \mbox{0.1810\,{\scriptsize (0.0033)}} & \mbox{0.4381\,{\scriptsize (0.0033)}} & \mbox{0.1822\,{\scriptsize (0.0034)}} & \mbox{0.1235\,{\scriptsize (0.0032)}} & \mbox{0.1305\,{\scriptsize (0.0034)}}\\

 &  &  & 4.00 & \textbf{\mbox{0.0506\,{\scriptsize (0.0014)}}} & \mbox{0.1761\,{\scriptsize (0.0031)}} & \mbox{0.7290\,{\scriptsize (0.0065)}} & \mbox{0.1761\,{\scriptsize (0.0031)}} & \mbox{0.0530\,{\scriptsize (0.0015)}} & \mbox{0.0535\,{\scriptsize (0.0014)}}\\

 &  & H & 0.25 & \mbox{0.1532\,{\scriptsize (0.0026)}} & \textbf{\mbox{0.1446\,{\scriptsize (0.0023)}}} & \mbox{0.1927\,{\scriptsize (0.0013)}} & \mbox{0.1489\,{\scriptsize (0.0023)}} & \mbox{0.1523\,{\scriptsize (0.0027)}} & \mbox{0.1461\,{\scriptsize (0.0024)}}\\

 &  &  & 1.00 & \textbf{\mbox{0.1326\,{\scriptsize (0.0031)}}} & \mbox{0.1698\,{\scriptsize (0.0030)}} & \mbox{0.4332\,{\scriptsize (0.0033)}} & \mbox{0.1703\,{\scriptsize (0.0030)}} & \mbox{0.1355\,{\scriptsize (0.0031)}} & \mbox{0.1355\,{\scriptsize (0.0029)}}\\

 &  &  & 4.00 & \textbf{\mbox{0.0826\,{\scriptsize (0.0023)}}} & \mbox{0.1744\,{\scriptsize (0.0031)}} & \mbox{0.7297\,{\scriptsize (0.0059)}} & \mbox{0.1744\,{\scriptsize (0.0031)}} & \mbox{0.0859\,{\scriptsize (0.0023)}} & \mbox{0.0933\,{\scriptsize (0.0024)}}\\

 &  & X & 1.00 & \textbf{\mbox{0.0948\,{\scriptsize (0.0021)}}} & \mbox{0.1255\,{\scriptsize (0.0024)}} & \mbox{0.4328\,{\scriptsize (0.0030)}} & \mbox{0.1256\,{\scriptsize (0.0024)}} & \mbox{0.0955\,{\scriptsize (0.0022)}} & \mbox{0.0955\,{\scriptsize (0.0020)}}\\

 & 0.3 & E & 0.25 & \mbox{0.1432\,{\scriptsize (0.0022)}} & \textbf{\mbox{0.1252\,{\scriptsize (0.0020)}}} & \mbox{0.1554\,{\scriptsize (0.0014)}} & \mbox{0.1279\,{\scriptsize (0.0019)}} & \mbox{0.1447\,{\scriptsize (0.0023)}} & \mbox{0.1324\,{\scriptsize (0.0020)}}\\

 &  &  & 1.00 & \mbox{0.1798\,{\scriptsize (0.0043)}} & \textbf{\mbox{0.1503\,{\scriptsize (0.0028)}}} & \mbox{0.3270\,{\scriptsize (0.0029)}} & \mbox{0.1537\,{\scriptsize (0.0031)}} & \mbox{0.1766\,{\scriptsize (0.0039)}} & \mbox{0.1840\,{\scriptsize (0.0035)}}\\

 &  &  & 4.00 & \textbf{\mbox{0.0554\,{\scriptsize (0.0015)}}} & \mbox{0.1521\,{\scriptsize (0.0031)}} & \mbox{0.5705\,{\scriptsize (0.0050)}} & \mbox{0.1521\,{\scriptsize (0.0031)}} & \mbox{0.0610\,{\scriptsize (0.0017)}} & \mbox{0.0636\,{\scriptsize (0.0017)}}\\

 &  & H & 0.25 & \mbox{0.1283\,{\scriptsize (0.0019)}} & \textbf{\mbox{0.1133\,{\scriptsize (0.0018)}}} & \mbox{0.1587\,{\scriptsize (0.0012)}} & \mbox{0.1175\,{\scriptsize (0.0019)}} & \mbox{0.1271\,{\scriptsize (0.0020)}} & \mbox{0.1182\,{\scriptsize (0.0017)}}\\

 &  &  & 1.00 & \mbox{0.1497\,{\scriptsize (0.0032)}} & \textbf{\mbox{0.1410\,{\scriptsize (0.0024)}}} & \mbox{0.3339\,{\scriptsize (0.0025)}} & \mbox{0.1426\,{\scriptsize (0.0024)}} & \mbox{0.1470\,{\scriptsize (0.0032)}} & \mbox{0.1478\,{\scriptsize (0.0030)}}\\

 &  &  & 4.00 & \mbox{0.1128\,{\scriptsize (0.0028)}} & \mbox{0.1558\,{\scriptsize (0.0031)}} & \mbox{0.5804\,{\scriptsize (0.0052)}} & \mbox{0.1558\,{\scriptsize (0.0031)}} & \textbf{\mbox{0.1090\,{\scriptsize (0.0028)}}} & \mbox{0.1129\,{\scriptsize (0.0025)}}\\

 & 0.6 & E & 0.25 & \mbox{0.1152\,{\scriptsize (0.0016)}} & \textbf{\mbox{0.0933\,{\scriptsize (0.0016)}}} & \mbox{0.1289\,{\scriptsize (0.0013)}} & \mbox{0.0958\,{\scriptsize (0.0016)}} & \mbox{0.1127\,{\scriptsize (0.0018)}} & \mbox{0.1017\,{\scriptsize (0.0015)}}\\

 &  &  & 1.00 & \mbox{0.2000\,{\scriptsize (0.0035)}} & \textbf{\mbox{0.1335\,{\scriptsize (0.0024)}}} & \mbox{0.2428\,{\scriptsize (0.0024)}} & \mbox{0.1355\,{\scriptsize (0.0025)}} & \mbox{0.2029\,{\scriptsize (0.0033)}} & \mbox{0.2011\,{\scriptsize (0.0029)}}\\

 &  &  & 4.00 & \textbf{\mbox{0.1106\,{\scriptsize (0.0044)}}} & \mbox{0.1403\,{\scriptsize (0.0027)}} & \mbox{0.3995\,{\scriptsize (0.0036)}} & \mbox{0.1409\,{\scriptsize (0.0027)}} & \mbox{0.1173\,{\scriptsize (0.0045)}} & \mbox{0.1155\,{\scriptsize (0.0042)}}\\

 &  & H & 0.25 & \mbox{0.1081\,{\scriptsize (0.0017)}} & \textbf{\mbox{0.0869\,{\scriptsize (0.0017)}}} & \mbox{0.1303\,{\scriptsize (0.0014)}} & \mbox{0.0892\,{\scriptsize (0.0017)}} & \mbox{0.1051\,{\scriptsize (0.0017)}} & \mbox{0.0989\,{\scriptsize (0.0017)}}\\

 &  &  & 1.00 & \mbox{0.1733\,{\scriptsize (0.0033)}} & \textbf{\mbox{0.1232\,{\scriptsize (0.0022)}}} & \mbox{0.2371\,{\scriptsize (0.0024)}} & \mbox{0.1258\,{\scriptsize (0.0023)}} & \mbox{0.1735\,{\scriptsize (0.0033)}} & \mbox{0.1707\,{\scriptsize (0.0029)}}\\

 &  &  & 4.00 & \textbf{\mbox{0.1372\,{\scriptsize (0.0038)}}} & \mbox{0.1429\,{\scriptsize (0.0029)}} & \mbox{0.4004\,{\scriptsize (0.0037)}} & \mbox{0.1435\,{\scriptsize (0.0030)}} & \mbox{0.1423\,{\scriptsize (0.0039)}} & \mbox{0.1403\,{\scriptsize (0.0038)}}\\

 & 0.9 & E & 0.25 & \mbox{0.0839\,{\scriptsize (0.0023)}} & \textbf{\mbox{0.0823\,{\scriptsize (0.0018)}}} & \mbox{0.1187\,{\scriptsize (0.0019)}} & \mbox{0.0858\,{\scriptsize (0.0019)}} & \mbox{0.0882\,{\scriptsize (0.0024)}} & \mbox{0.0835\,{\scriptsize (0.0019)}}\\

 &  &  & 1.00 & \mbox{0.1456\,{\scriptsize (0.0030)}} & \textbf{\mbox{0.1086\,{\scriptsize (0.0018)}}} & \mbox{0.1777\,{\scriptsize (0.0021)}} & \mbox{0.1099\,{\scriptsize (0.0020)}} & \mbox{0.1496\,{\scriptsize (0.0028)}} & \mbox{0.1491\,{\scriptsize (0.0027)}}\\

 &  &  & 4.00 & \mbox{0.1717\,{\scriptsize (0.0045)}} & \textbf{\mbox{0.1394\,{\scriptsize (0.0023)}}} & \mbox{0.2730\,{\scriptsize (0.0029)}} & \mbox{0.1409\,{\scriptsize (0.0023)}} & \mbox{0.1989\,{\scriptsize (0.0044)}} & \mbox{0.1892\,{\scriptsize (0.0040)}}\\

 &  & H & 0.25 & \mbox{0.0698\,{\scriptsize (0.0020)}} & \textbf{\mbox{0.0685\,{\scriptsize (0.0015)}}} & \mbox{0.1020\,{\scriptsize (0.0016)}} & \mbox{0.0704\,{\scriptsize (0.0015)}} & \mbox{0.0715\,{\scriptsize (0.0020)}} & \mbox{0.0709\,{\scriptsize (0.0020)}}\\

 &  &  & 1.00 & \mbox{0.1196\,{\scriptsize (0.0025)}} & \textbf{\mbox{0.0997\,{\scriptsize (0.0020)}}} & \mbox{0.1613\,{\scriptsize (0.0022)}} & \mbox{0.1026\,{\scriptsize (0.0022)}} & \mbox{0.1186\,{\scriptsize (0.0026)}} & \mbox{0.1191\,{\scriptsize (0.0025)}}\\

 &  &  & 4.00 & \mbox{0.1505\,{\scriptsize (0.0035)}} & \textbf{\mbox{0.1318\,{\scriptsize (0.0021)}}} & \mbox{0.2389\,{\scriptsize (0.0025)}} & \mbox{0.1334\,{\scriptsize (0.0022)}} & \mbox{0.1667\,{\scriptsize (0.0040)}} & \mbox{0.1617\,{\scriptsize (0.0038)}}\\

 &  & X & 1.00 & \mbox{0.0685\,{\scriptsize (0.0017)}} & \textbf{\mbox{0.0494\,{\scriptsize (0.0013)}}} & \mbox{0.1315\,{\scriptsize (0.0018)}} & \mbox{0.0514\,{\scriptsize (0.0014)}} & \mbox{0.0704\,{\scriptsize (0.0018)}} & \mbox{0.0712\,{\scriptsize (0.0017)}}\\
\cmidrule{1-10}
500 & 0.0 & E & 0.25 & \mbox{0.2133\,{\scriptsize (0.0020)}} & \mbox{0.2107\,{\scriptsize (0.0022)}} & \mbox{0.2414\,{\scriptsize (0.0008)}} & \mbox{0.2137\,{\scriptsize (0.0021)}} & \mbox{0.2164\,{\scriptsize (0.0021)}} & \textbf{\mbox{0.2096\,{\scriptsize (0.0021)}}}\\

 &  &  & 1.00 & \textbf{\mbox{0.1810\,{\scriptsize (0.0051)}}} & \mbox{0.2769\,{\scriptsize (0.0046)}} & \mbox{0.8097\,{\scriptsize (0.0026)}} & \mbox{0.2778\,{\scriptsize (0.0047)}} & \mbox{0.1852\,{\scriptsize (0.0048)}} & \mbox{0.2132\,{\scriptsize (0.0050)}}\\

 &  &  & 4.00 & \textbf{\mbox{0.0529\,{\scriptsize (0.0013)}}} & \mbox{0.2806\,{\scriptsize (0.0048)}} & \mbox{2.6414\,{\scriptsize (0.0087)}} & \mbox{0.2806\,{\scriptsize (0.0048)}} & \mbox{0.0578\,{\scriptsize (0.0015)}} & \mbox{0.0629\,{\scriptsize (0.0015)}}\\

 &  & H & 0.25 & \textbf{\mbox{0.1800\,{\scriptsize (0.0026)}}} & \mbox{0.1823\,{\scriptsize (0.0024)}} & \mbox{0.2425\,{\scriptsize (0.0011)}} & \mbox{0.1868\,{\scriptsize (0.0025)}} & \mbox{0.1817\,{\scriptsize (0.0026)}} & \mbox{0.1805\,{\scriptsize (0.0024)}}\\

 &  &  & 1.00 & \textbf{\mbox{0.1904\,{\scriptsize (0.0039)}}} & \mbox{0.2613\,{\scriptsize (0.0039)}} & \mbox{0.8083\,{\scriptsize (0.0023)}} & \mbox{0.2614\,{\scriptsize (0.0039)}} & \mbox{0.1921\,{\scriptsize (0.0038)}} & \mbox{0.2051\,{\scriptsize (0.0038)}}\\

 &  &  & 4.00 & \mbox{0.1227\,{\scriptsize (0.0033)}} & \mbox{0.2840\,{\scriptsize (0.0044)}} & \mbox{2.6320\,{\scriptsize (0.0080)}} & \mbox{0.2840\,{\scriptsize (0.0044)}} & \textbf{\mbox{0.1207\,{\scriptsize (0.0034)}}} & \mbox{0.1395\,{\scriptsize (0.0031)}}\\

 &  & X & 1.00 & \textbf{\mbox{0.1229\,{\scriptsize (0.0028)}}} & \mbox{0.1813\,{\scriptsize (0.0031)}} & \mbox{0.8088\,{\scriptsize (0.0023)}} & \mbox{0.1813\,{\scriptsize (0.0031)}} & \mbox{0.1238\,{\scriptsize (0.0030)}} & \mbox{0.1306\,{\scriptsize (0.0027)}}\\

 & 0.3 & E & 0.25 & \mbox{0.1707\,{\scriptsize (0.0021)}} & \mbox{0.1661\,{\scriptsize (0.0021)}} & \mbox{0.2249\,{\scriptsize (0.0011)}} & \mbox{0.1699\,{\scriptsize (0.0024)}} & \mbox{0.1744\,{\scriptsize (0.0022)}} & \textbf{\mbox{0.1657\,{\scriptsize (0.0021)}}}\\

 &  &  & 1.00 & \mbox{0.2699\,{\scriptsize (0.0040)}} & \textbf{\mbox{0.2257\,{\scriptsize (0.0034)}}} & \mbox{0.6791\,{\scriptsize (0.0026)}} & \mbox{0.2280\,{\scriptsize (0.0035)}} & \mbox{0.2506\,{\scriptsize (0.0042)}} & \mbox{0.2635\,{\scriptsize (0.0036)}}\\

 &  &  & 4.00 & \textbf{\mbox{0.0696\,{\scriptsize (0.0021)}}} & \mbox{0.2251\,{\scriptsize (0.0041)}} & \mbox{1.9458\,{\scriptsize (0.0080)}} & \mbox{0.2251\,{\scriptsize (0.0041)}} & \mbox{0.0812\,{\scriptsize (0.0023)}} & \mbox{0.0893\,{\scriptsize (0.0022)}}\\

 &  & H & 0.25 & \mbox{0.1463\,{\scriptsize (0.0021)}} & \mbox{0.1453\,{\scriptsize (0.0023)}} & \mbox{0.2252\,{\scriptsize (0.0010)}} & \mbox{0.1500\,{\scriptsize (0.0025)}} & \mbox{0.1474\,{\scriptsize (0.0022)}} & \textbf{\mbox{0.1442\,{\scriptsize (0.0022)}}}\\

 &  &  & 1.00 & \mbox{0.2184\,{\scriptsize (0.0039)}} & \textbf{\mbox{0.2149\,{\scriptsize (0.0034)}}} & \mbox{0.6864\,{\scriptsize (0.0028)}} & \mbox{0.2159\,{\scriptsize (0.0035)}} & \mbox{0.2152\,{\scriptsize (0.0039)}} & \mbox{0.2215\,{\scriptsize (0.0035)}}\\

 &  &  & 4.00 & \mbox{0.1401\,{\scriptsize (0.0029)}} & \mbox{0.2259\,{\scriptsize (0.0036)}} & \mbox{1.9771\,{\scriptsize (0.0086)}} & \mbox{0.2259\,{\scriptsize (0.0036)}} & \textbf{\mbox{0.1331\,{\scriptsize (0.0032)}}} & \mbox{0.1427\,{\scriptsize (0.0029)}}\\

 & 0.6 & E & 0.25 & \mbox{0.1280\,{\scriptsize (0.0018)}} & \textbf{\mbox{0.1229\,{\scriptsize (0.0020)}}} & \mbox{0.2130\,{\scriptsize (0.0011)}} & \mbox{0.1267\,{\scriptsize (0.0021)}} & \mbox{0.1287\,{\scriptsize (0.0020)}} & \mbox{0.1236\,{\scriptsize (0.0019)}}\\

 &  &  & 1.00 & \mbox{0.2406\,{\scriptsize (0.0035)}} & \textbf{\mbox{0.1744\,{\scriptsize (0.0027)}}} & \mbox{0.5730\,{\scriptsize (0.0030)}} & \mbox{0.1770\,{\scriptsize (0.0027)}} & \mbox{0.2408\,{\scriptsize (0.0033)}} & \mbox{0.2471\,{\scriptsize (0.0031)}}\\

 &  &  & 4.00 & \textbf{\mbox{0.2057\,{\scriptsize (0.0076)}}} & \mbox{0.2111\,{\scriptsize (0.0038)}} & \mbox{1.3956\,{\scriptsize (0.0083)}} & \mbox{0.2117\,{\scriptsize (0.0038)}} & \mbox{0.2357\,{\scriptsize (0.0073)}} & \mbox{0.2334\,{\scriptsize (0.0069)}}\\

 &  & H & 0.25 & \mbox{0.1156\,{\scriptsize (0.0017)}} & \textbf{\mbox{0.1079\,{\scriptsize (0.0020)}}} & \mbox{0.2126\,{\scriptsize (0.0012)}} & \mbox{0.1117\,{\scriptsize (0.0023)}} & \mbox{0.1153\,{\scriptsize (0.0018)}} & \mbox{0.1100\,{\scriptsize (0.0017)}}\\

 &  &  & 1.00 & \mbox{0.2152\,{\scriptsize (0.0031)}} & \textbf{\mbox{0.1673\,{\scriptsize (0.0027)}}} & \mbox{0.5741\,{\scriptsize (0.0029)}} & \mbox{0.1697\,{\scriptsize (0.0028)}} & \mbox{0.2145\,{\scriptsize (0.0032)}} & \mbox{0.2212\,{\scriptsize (0.0030)}}\\

 &  &  & 4.00 & \mbox{0.2129\,{\scriptsize (0.0049)}} & \textbf{\mbox{0.2110\,{\scriptsize (0.0035)}}} & \mbox{1.3737\,{\scriptsize (0.0077)}} & \mbox{0.2115\,{\scriptsize (0.0035)}} & \mbox{0.2295\,{\scriptsize (0.0050)}} & \mbox{0.2342\,{\scriptsize (0.0047)}}\\

 & 0.9 & E & 0.25 & \textbf{\mbox{0.0942\,{\scriptsize (0.0028)}}} & \mbox{0.1236\,{\scriptsize (0.0025)}} & \mbox{0.2440\,{\scriptsize (0.0012)}} & \mbox{0.1263\,{\scriptsize (0.0026)}} & \mbox{0.0987\,{\scriptsize (0.0026)}} & \mbox{0.1042\,{\scriptsize (0.0026)}}\\

 &  &  & 1.00 & \mbox{0.1829\,{\scriptsize (0.0032)}} & \textbf{\mbox{0.1597\,{\scriptsize (0.0027)}}} & \mbox{0.5201\,{\scriptsize (0.0041)}} & \mbox{0.1625\,{\scriptsize (0.0028)}} & \mbox{0.1915\,{\scriptsize (0.0030)}} & \mbox{0.1911\,{\scriptsize (0.0027)}}\\

 &  &  & 4.00 & \mbox{0.2539\,{\scriptsize (0.0046)}} & \textbf{\mbox{0.2153\,{\scriptsize (0.0033)}}} & \mbox{0.9810\,{\scriptsize (0.0076)}} & \mbox{0.2173\,{\scriptsize (0.0034)}} & \mbox{0.2820\,{\scriptsize (0.0045)}} & \mbox{0.2757\,{\scriptsize (0.0044)}}\\

 &  & H & 0.25 & \textbf{\mbox{0.0713\,{\scriptsize (0.0012)}}} & \mbox{0.0837\,{\scriptsize (0.0016)}} & \mbox{0.2143\,{\scriptsize (0.0016)}} & \mbox{0.0856\,{\scriptsize (0.0018)}} & \mbox{0.0752\,{\scriptsize (0.0017)}} & \mbox{0.0761\,{\scriptsize (0.0015)}}\\

 &  &  & 1.00 & \textbf{\mbox{0.1370\,{\scriptsize (0.0029)}}} & \mbox{0.1388\,{\scriptsize (0.0024)}} & \mbox{0.4442\,{\scriptsize (0.0033)}} & \mbox{0.1419\,{\scriptsize (0.0024)}} & \mbox{0.1404\,{\scriptsize (0.0030)}} & \mbox{0.1474\,{\scriptsize (0.0029)}}\\

 &  &  & 4.00 & \mbox{0.2075\,{\scriptsize (0.0042)}} & \textbf{\mbox{0.1917\,{\scriptsize (0.0028)}}} & \mbox{0.8040\,{\scriptsize (0.0059)}} & \mbox{0.1933\,{\scriptsize (0.0028)}} & \mbox{0.2352\,{\scriptsize (0.0041)}} & \mbox{0.2318\,{\scriptsize (0.0042)}}\\

 &  & X & 1.00 & \mbox{0.0811\,{\scriptsize (0.0015)}} & \textbf{\mbox{0.0623\,{\scriptsize (0.0016)}}} & \mbox{0.3575\,{\scriptsize (0.0028)}} & \mbox{0.0644\,{\scriptsize (0.0017)}} & \mbox{0.0871\,{\scriptsize (0.0019)}} & \mbox{0.0869\,{\scriptsize (0.0017)}}\\
\bottomrule
\end{tabular}
}
\end{table}

Table~\ref{tab:supp_excess_risk} isolates prediction risk arising from
estimation. Conditional on a fitted model, expected NMSE equals one plus
ER; observed NMSE additionally varies with the independent test sample.
For equal signals at $p=500$, $\rho=0$, and SNR 4, ER is 0.0529 for
Lambert, 0.0578 for MCP, 0.0629 for SCAD, 0.2806 for both LASSO and
ENET, and 2.6414 for ridge. Thus the favorable Lambert comparison in
Table~\ref{tab:nmse_combined} also appears without test-sample noise.

At $\rho=0.6$, Lambert remains better than MCP and SCAD (0.2057 versus
0.2357 and 0.2334). At $\rho=0.9$, LASSO instead improves on Lambert
(0.2153 versus 0.2539), although Lambert still improves on both nonconvex
comparators. Moderate heterogeneity at $\rho=0$ also favors MCP slightly
(0.1207 versus 0.1227). Unlike CE, ER weights coefficient errors by the
predictor covariance and scales by noise variance. This explains why a
convex estimator can have much smaller coefficient error without the same
advantage in prediction. Lambert's favorable regime therefore depends on
the signal structure and the assessment criterion.

No incomplete CV searches were recorded for Lambert or ridge. Restricting
the comparisons to pairs with complete searches preserved the main
prediction and variable-selection patterns (Table~\ref{tab:fixed_sensitivity}).

\begin{table}[!htbp]
\centering\spacingset{1}\scriptsize
\setlength{\tabcolsep}{2pt}
\renewcommand{\arraystretch}{1.15}
\caption{Sensitivity of prediction and support-recovery comparisons to incomplete cross-validation searches. Differences are Lambert minus comparator, averaged over included pairs with complete searches for both methods. Negative NMSE and positive MCC differences favor Lambert. Main and stress results are separate; exclusion changes the scenario composition of the retained pairs.}
\label{tab:fixed_sensitivity}

\begin{tabular}{llrrrr}
\toprule
Scope & Comparator & Included & Excluded & NMSE difference & MCC difference\\
\midrule
Main & LASSO & 13725 & 675 & -0.01481 & 0.05950\\

 & ENET & 13725 & 675 & -0.01665 & 0.07888\\

 & MCP & 14330 & 70 & -0.00410 & -0.02227\\

 & SCAD & 14340 & 60 & -0.00472 & 0.05863\\
\cmidrule{1-6}
Stress & LASSO & 1192 & 8 & -0.01269 & -0.01587\\

 & ENET & 1192 & 8 & -0.01369 & -0.01448\\

 & MCP & 1187 & 13 & -0.00213 & -0.02216\\

 & SCAD & 1194 & 6 & -0.00416 & 0.04840\\
\bottomrule
\end{tabular}

\end{table}

Table~\ref{tab:fixed_sensitivity} retains 14,330 main-design pairs for MCP
and 14,340 for SCAD. The Lambert-minus-MCP differences are negative for both
NMSE ($-0.00410$) and MCC ($-0.02227$), preserving the contrast between
Lambert's lower prediction error and MCP's better support recovery. Against
SCAD, Lambert has both lower NMSE ($-0.00472$) and higher MCC (0.05863).
Against LASSO and ENET, 13,725 main-design pairs remain for each comparison.
The respective NMSE differences are $-0.01481$ and $-0.01665$, with MCC
gains of 0.05950 and 0.07888. Thus the conditional averages favor Lambert
over these convex sparse comparators on both measures, while the MCP
comparison still favors different methods for prediction and selection.
These averages do not imply an advantage in every constituent scenario.
Ridge has no incomplete searches, so its full-sample comparison in
Tables~\ref{tab:nmse_combined} and~\ref{tab:mcc_combined} is unchanged.

The stress scenarios give a different selection pattern. Lambert retains
lower average NMSE than each comparator, but its MCC differences are negative
against LASSO, ENET, and MCP ($-0.01587$, $-0.01448$, and $-0.02216$,
respectively). Against SCAD, NMSE improves by 0.00416 and MCC by 0.04840.
Thus the prediction--selection tradeoff persists when
incomplete searches are excluded. However, the included pairs differ across
comparators, and exclusion can change the mix of replications and scenarios.
This conditional analysis supports the direction of the main comparisons
without replacing the full results or showing that incomplete paths have
no effect on individual comparisons.

Taken together, Tables~\ref{tab:nmse_combined}--\ref{tab:supp_excess_risk}
identify a regime-dependent role for Lambert. The controlled release of
shrinkage in Section~\ref{sec:retentiondesign} gives a continuous entry rule
and a finite identity tail. Its earlier identity tail than SCAD, at the
fixed shapes used here, is compatible with reduced shrinkage of strong
signals: at a common penalty level, Lambert leaves scalar scores above
$e\lambda$ unchanged, while SCAD reaches identity at $3.7\lambda$
(Figure~\ref{fig:penalties}). LASSO continues to shrink large scores, which
can increase bias in strong-signal settings, but its simpler shrinkage can
be effective when signals are weak. Ridge and ENET instead stabilize
correlated coefficient estimates through quadratic penalization
\citep{tibshirani1996,zouhastie2005}. Their CE advantages show why this
stabilization remains useful even when it does not minimize prediction loss.
MCP also has a finite identity tail and retains an advantage for support
recovery in most main scenarios. These scalar and penalty-level explanations
are consistent with the observed patterns; they do not isolate their cause,
because CV selects different penalty levels and the multivariate fit also
depends on predictor dependence.

\section{Real-data application: prediction of aquatic toxicity}
\label{sec:qsar}

We downloaded the QSAR aquatic toxicity data from the UCI Machine Learning
Repository (dataset 505) \citep{qsar2014}. The data contain $n=546$
chemicals and $p=8$ molecular descriptors from the quantitative
structure--activity relationship study of \citet{cassotti2014aquatic}.
The response is the negative logarithm of the molar concentration causing
50\% mortality in \emph{Daphnia magna} after 48 hours; higher values
therefore indicate greater toxicity.

The application asks whether a compact descriptor model can preserve
predictive accuracy and reproduce its variable selection under changes in
the training sample. For this use, a small reduction in prediction error is
most informative when considered alongside the number and consistency of
retained descriptors. The descriptors summarize molecular structure, and
their selection describes the fitted predictive representation; it does
not identify causal mechanisms of toxicity.

The supplied descriptors are TPSA(Tot), SAacc, H-050, MLOGP, RDCHI,
GATS1p, nN, and C-040. We treated these as fixed inputs to a common linear
regression model. All observations were retained, with no missing values,
imputation, or further response transformation.

\subsection{Evaluation protocol}
Lambert, ordinary least squares (OLS), LASSO, ridge, ENET, MCP, and SCAD
used identical observations and CV partitions, with tuning as in
Section~\ref{sec:implementation}. ENET additionally selected its mixing
parameter from $\{0,0.25,0.50,0.75,1\}$.

We used 100 repetitions of nested five-fold cross-validation. Each random
outer partition gave training sets of 436 or 437 observations and test sets
of 110 or 109, approximately an 80:20 split. Five-fold inner cross-validation
selected penalties using only the outer training set. The selected model was
fitted to that entire training set and evaluated on its held-out fold, giving
each observation one out-of-sample prediction per repetition.

Equations~\eqref{eq:lambdapath} and~\eqref{eq:cvscore} define the path
and selection rule. Predictor centering and root-mean-square scaling were
learned within each training split and applied to its validation or test
observations, excluding outer-test information from preprocessing and tuning.
The intercept was unpenalized; the response was not variance-standardized.

Let $I_{rf}$ denote the outer test set in repetition $r\in\{1,\ldots,100\}$
and fold $f\in\{1,\ldots,5\}$, and let $|I_{rf}|$ be its number of observations.
For observation $i$, $y_i$ is the observed response and
$\widehat y_{i,m}^{(r,f)}$ is the prediction from method $m$, trained and tuned
without observations in $I_{rf}$. We calculated the repetition-level score as
\begin{equation}
 R_{rm}=\frac{1}{5}\sum_{f=1}^{5}
 \left\{\frac{1}{|I_{rf}|}\sum_{i\in I_{rf}}
 \left(y_i-\widehat y_{i,m}^{(r,f)}\right)^2\right\}^{1/2}.
 \label{eq:qsar_rmse}
\end{equation}
Equation~\eqref{eq:qsar_rmse} weights the five fold RMSEs equally; we then
average the 100 values $R_{rm}$, rather than pool all held-out predictions.
MAE and selected size follow the same convention. We also report the sample
standard deviation (SD) across repetitions. For a generic repetition-level
score $V_{rm}$, write
$\bar V_m=100^{-1}\sum_{r=1}^{100}V_{rm}$; the reported SD is
\[
 s_m=\left\{\frac{1}{99}\sum_{r=1}^{100}(V_{rm}-\bar V_m)^2\right\}^{1/2}.
\]
This SD describes partition sensitivity, not a standard error of population
generalization performance; overlapping training samples require accounting
for dependence \citep{bengiograndvalet2004}. Stability is pooled and has no
repetition-level SD. Prediction errors and their SDs retain the response scale.

RMSE emphasizes large errors; MAE gives extreme residuals less weight.
Both assess prediction for unseen chemicals. Selected size describes
complexity and descriptor requirements: smaller models are useful only
with adequate accuracy. Stability measures reproducibility under changes
in training samples, without establishing biological correctness.

\subsection{Predictive accuracy and selection stability}
Table~\ref{tab:qsar_prediction} reports prediction, model size, and stability
for the seven methods on the same repeated partitions.

A predictor was counted as selected when the absolute original-scale
coefficient exceeded $10^{-8}$; this reporting threshold changed neither
coefficients nor predictions. Selection stability was summarized across the
500 outer-training fits using the chance-adjusted measure of
\citet{nogueira2018}. This measure describes the reproducibility of selected
subsets, not the correctness of variable selection. Because the true active
predictors are unknown, support-recovery MCC was not calculated.

\begin{table}[!htbp]
\centering\spacingset{1}\footnotesize
\setlength{\tabcolsep}{3pt}
\caption{Predictive accuracy, model sparsity, and selection stability in
the QSAR aquatic-toxicity application under repeated nested cross-validation.
Entries are means (SDs) for RMSE, MAE, and selected size, and pooled
Nogueira estimates for stability. Bold identifies the lowest unrounded error
means and highest finite stability; a dash denotes selection of all
predictors in every fit.}
\label{tab:qsar_prediction}

\begin{tabular}{>{}lrrrr}
\toprule
Method & RMSE & MAE & Selected size & Stability\\
\midrule
Lambert & \textbf{1.212844 (0.007284)} & \textbf{0.920094 (0.005055)} & 6.114 (0.161) & \textbf{0.9253}\\
LASSO & 1.216413 (0.007815) & 0.924524 (0.005349) & 7.092 (0.334) & 0.3988\\
Ridge & 1.217195 (0.007702) & 0.924046 (0.005421) & 8.000 (0.000) & --\\
ENET & 1.216673 (0.007750) & 0.924802 (0.005377) & 7.178 (0.329) & 0.3575\\
MCP & 1.213165 (0.007394) & 0.920320 (0.005207) & 6.126 (0.165) & 0.9151\\
SCAD & 1.213123 (0.007063) & 0.920266 (0.004962) & 6.172 (0.171) & 0.8883\\
OLS & 1.216553 (0.007574) & 0.921051 (0.005379) & 8.000 (0.000) & --\\
\bottomrule
\end{tabular}

\end{table}

Table~\ref{tab:qsar_prediction} places Lambert first in average RMSE and
MAE, with values of 1.212844 and 0.920094. Its RMSE is below MCP's
1.213165 and SCAD's 1.213123 by only 0.000321 and 0.000279, or about
0.026\% and 0.023\%. The difference from OLS is larger but still modest:
0.003709, approximately 0.305\%. LASSO, ridge, and ENET have RMSE between
1.216413 and 1.217195. The MAE comparison gives the same leading method,
with MCP and SCAD again close to Lambert. These results indicate competitive
prediction with small average gains rather than a practically large
improvement over the other nonconvex penalties.

The repeated partitions help assess how consistently these differences
occur. Lambert has lower repetition-level RMSE than OLS in 99 of 100
partitions, compared with 91 against LASSO, 99 against ridge, 93 against
ENET, 49 against MCP, and 56 against SCAD. Its MAE is lower than all three
convex penalized comparators in every repetition. Its best average RMSE
therefore does not correspond to a consistent advantage over
MCP across partitions. The RMSE SDs are also similar: 0.007284 for Lambert,
0.007394 for MCP, and 0.007063 for SCAD. These counts and SDs describe
sensitivity to the chosen partitions; the overlapping samples do not
provide independent trials for a significance test.

Model size and stability show a clearer distinction from the convex sparse
comparators. Lambert selects 6.114 descriptors on average, close to MCP
(6.126) and SCAD (6.172), but fewer than LASSO (7.092) and ENET (7.178).
OLS and ridge retain all eight descriptors. Lambert's stability of 0.9253
exceeds MCP's 0.9151 and SCAD's 0.8883, while LASSO and ENET yield 0.3988
and 0.3575. The higher stability accompanies similar or slightly smaller
models, without a loss of average predictive accuracy. Against LASSO and
ENET, the practical distinction is a more reproducible subset with roughly
one fewer descriptor. Against MCP and SCAD, model sizes and errors are
already close, leaving the higher estimated stability as the clearer
descriptive distinction. These comparisons use Table~\ref{tab:qsar_prediction}
jointly, because neither sparsity nor stability alone establishes a useful
predictive model.

Six descriptors occur in every Lambert fit. H-050 and C-040 are selected
in only 6.4\% and 5.0\% of fits, compared with 46.6\% and 62.6\% for
LASSO and 51.4\% and 66.4\% for ENET. These concentrated selection frequencies
explain Lambert's higher stability. The six consistently retained descriptors
are TPSA(Tot), SAacc, MLOGP, RDCHI, GATS1p, and nN. In this dataset, the
result is a recurring core representation with occasional additions, which
helps distinguish stable predictive structure from partition-sensitive
inclusions. Rare selection of H-050 and C-040 does not establish their
biological irrelevance. The practical finding is therefore the combination
of compactness, repeatable selection, and competitive prediction; the small
error differences from MCP and SCAD give little basis for a claim of
substantially better toxicity prediction.

\subsection{Regularization paths and model size}
\begin{figure}[!htbp]
\centering\spacingset{1}
\includegraphics[width=\textwidth,height=0.70\textheight,keepaspectratio]{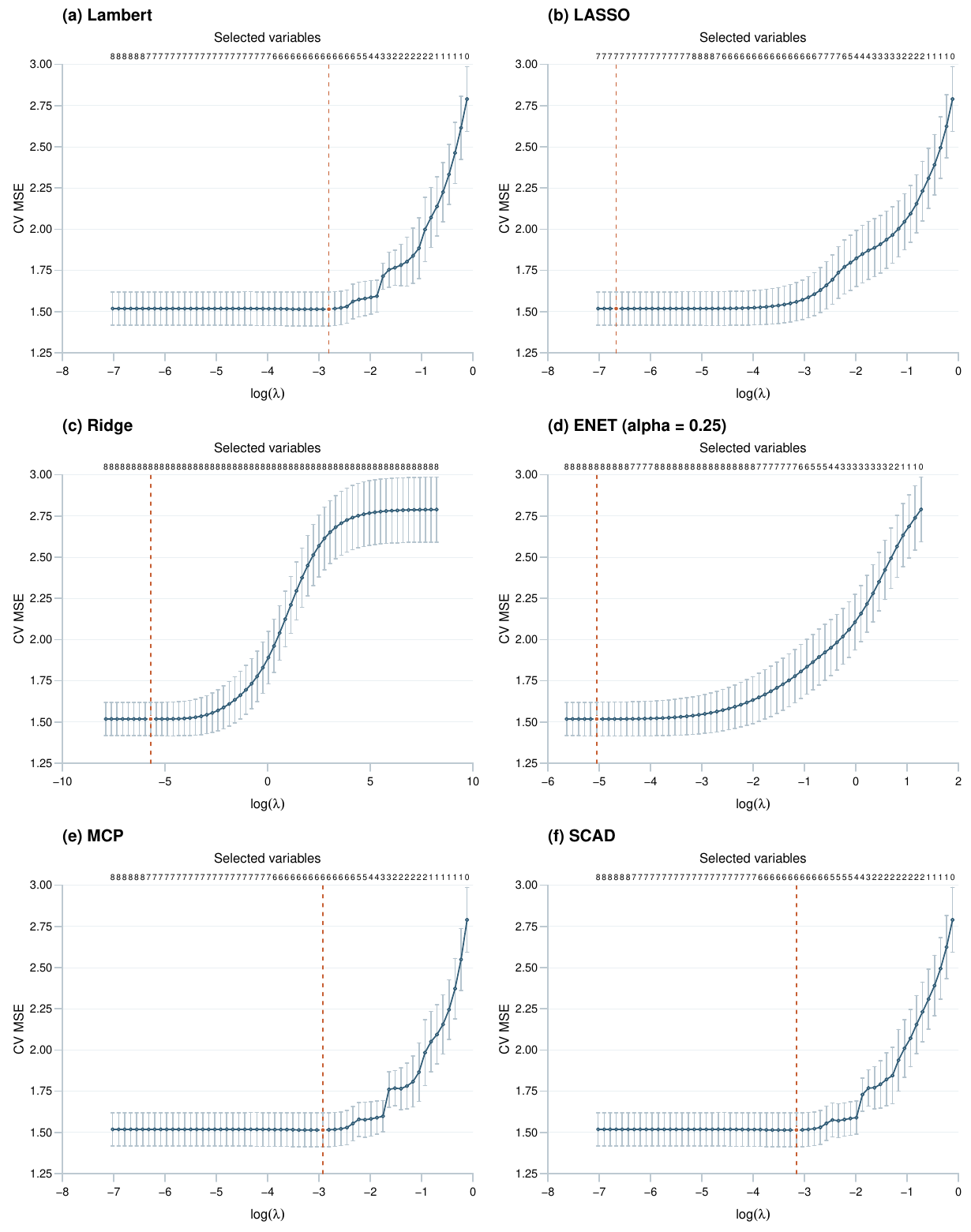}
\caption{Five-fold cross-validation paths and model size for the QSAR
application. Open circles and lines show mean validation MSE; error bars extend one
fold-based standard error on either side. The filled point and dashed line mark the selected
penalty. Upper labels count selected predictors in full-data fits, excluding
the intercept. Penalties increase from left to right; horizontal ranges follow
each method's tuning grid. The ENET curve uses the selected $\alpha=0.25$.
The selected $\lambda_{\min}$ values for panels (a)--(f) are, respectively,
0.0603045, 0.00126585, 0.00333188, 0.00639939, 0.0536416, 0.0424431.
These selections refer to this single partition, not averages across the
repeated nested-CV analysis.}
\label{fig:qsar_cv_paths}
\end{figure}

Figure~\ref{fig:qsar_cv_paths} illustrates how the tuning criterion and model
size change along each regularization path.

For this illustration, we applied
five-fold CV to all 546 observations, using the first recorded outer partition
as a common division for the six penalized methods. Preprocessing was learned
within each training fold, and the tuning grids and selection rules were
unchanged. The ENET panel displays the path at its jointly selected mixing
parameter, $\alpha=0.25$, rather than an envelope of minima across mixing
parameters. Horizontal coordinates are full-data penalty values; the CV
candidates remain aligned by grid position under the training-fold scaling
in Section~\ref{sec:implementation}. We denote the full-data penalty
corresponding to the minimum-CV choice by $\lambda_{\min}$; its value for
each method is given in the figure caption. Penalty magnitudes are
method-specific and do not represent equivalent shrinkage across methods.

The nonconvex paths in Figure~\ref{fig:qsar_cv_paths} reach a nearly flat
low-error region while retaining six descriptors at the selected penalty.
Lambert, MCP, and SCAD have effectively identical minimum CV MSEs
(1.515170 after rounding). LASSO selects seven descriptors, whereas ridge
and ENET select all eight, with minimum CV MSEs between 1.518530 and
1.518926. Ridge retains all eight throughout its path: its regularization
changes coefficient magnitudes without producing sparse models. The
upper labels therefore complement the error curves by showing how much
model complexity accompanies a given tuning choice.

This single-partition illustration is consistent with the compact nonconvex
models in Table~\ref{tab:qsar_prediction}, but does not establish a
predictive advantage among Lambert, MCP, and SCAD. Because these CV scores
also determine the tuning choices, the repeated nested-CV results in
Table~\ref{tab:qsar_prediction} remain the basis for the predictive comparison.
At each candidate, the standard error is the sample standard deviation of
the five fold MSEs divided by $\sqrt{5}$. The bars describe variation among
validation folds; because training samples overlap, they do not provide
a significance test for differences between methods.

\FloatBarrier
\section{Discussion and limitations}
\label{sec:discussion}

Lambert formalizes a choice about how a variable moves from exclusion to
full retention. The constraint in Equation~\eqref{eq:mainretentionbudget}
controls changes in the retained fraction on a logarithmic score scale;
Equation~\eqref{eq:mainthreshold} maximizes retention within that class.
This gives the penalty a statistical motivation and an explicit construction,
while Section~\ref{sec:audit} identifies its structural differences from
related penalties. Its practical simplicity follows from fixing the shape
and selecting one penalty level. Proposition~\ref{prop:mainprox} ensures a
unique scalar update, Theorem~\ref{thm:globalunique} identifies a sufficient
condition for multivariate uniqueness, and Theorem~\ref{thm:cd} establishes
descent and conditional convergence. Theorem~\ref{thm:mainoracle} adds
uniform oracle equivalence for global minimizers under its asymptotic
conditions. Together these results explain how the design leads to an
implementable estimator and where its guarantees apply.

The simulations connect this construction to distinct statistical aims.
Table~\ref{tab:nmse_combined} identifies strong equal signals as a favorable
prediction regime, including settings with moderate predictor correlation.
The finite identity tail is consistent with preserving sufficiently strong
scores, and the logarithmic transition determines how intermediate scores
are retained. This interpretation concerns the complete CV-tuned procedure;
MCP and SCAD also have identity tails, and the study does not attribute the
observed gains to that property alone. For support recovery,
Table~\ref{tab:mcc_combined} generally favors Lambert over SCAD and MCP over
Lambert. LASSO remains a strong prediction comparator at low SNR, while
ridge and ENET can improve coefficient reconstruction under strong
correlation. Tables~\ref{tab:supp_coefficient_error} and
\ref{tab:supp_excess_risk} clarify this distinction: correlated coefficient
errors can offset in the linear predictor. The appropriate method therefore
depends on whether the principal aim is prediction, variable discovery, or
estimation of individual effects. The complete-search comparison in
Table~\ref{tab:fixed_sensitivity} preserves the main patterns.

QSAR illustrates the practical value of considering these aims together.
Table~\ref{tab:qsar_prediction} gives Lambert the smallest average errors,
with only minor differences from MCP and SCAD, and the highest estimated
stability among the sparse methods. Six descriptors appear in every Lambert
fit, while two are included only occasionally. This recurring subset offers
a compact predictive representation whose composition changes little across
training partitions. Figure~\ref{fig:qsar_cv_paths} complements that result
by showing low-error nonconvex fits with six descriptors, whereas the
selected convex fits retain larger models. The application supports using
Lambert when reproducibility of a small predictor set accompanies the goal
of accurate prediction. The package in Section~\ref{sec:reproducibility}
provides the path, CV, prediction, and diagnostic tools needed to use that
procedure on another dataset.

Two limitations define the scope of these findings. The oracle result
assumes fixed dimension and global minimizers under the stated design and
tuning conditions; it does not establish the same guarantee for
CV-selected stationary solutions in increasing dimension. Empirical
conclusions concern the evaluated Gaussian sparse designs and the QSAR
application; broader covariance structures, error distributions, and
applications remain directions for further evaluation.

\section{Conclusion}
\label{sec:conclusion}

The Lambert penalty derives a sparse regression rule from a constraint on
the release of shrinkage. The resulting logarithmic transition joins exact
zeros to full retention, and its induced penalty admits a closed-form
coordinate update. The curvature, convergence, and oracle results connect
this construction to objective geometry, computation, and estimation under
stated assumptions.

The evaluated settings identify where the method is useful. Lambert favors
prediction for strong equal signals (Table~\ref{tab:nmse_combined}) and
usually improves support recovery over SCAD, while MCP is often stronger
on that criterion (Table~\ref{tab:mcc_combined}). Coefficient accuracy and
prediction risk can favor different methods, particularly under strong
correlation (Tables~\ref{tab:supp_coefficient_error} and
\ref{tab:supp_excess_risk}). In QSAR, the useful outcome is competitive
prediction with a compact and stable descriptor set
(Table~\ref{tab:qsar_prediction}). With one CV-selected penalty level and
an accompanying R implementation, Lambert offers a practical alternative
when both prediction and reproducible sparse representation matter.
Extensions to increasing-dimensional theory and broader applications remain
natural directions for further work.

\section{Computational implementation}
\label{sec:reproducibility}
LASSO, ridge, and ENET were fitted using the R package \texttt{glmnet},
while MCP and SCAD were fitted using \texttt{ncvreg}.
The Lambert method is implemented in the R package \texttt{lambertReg},
available at \url{https://github.com/byuzbasi/lambertReg}.

\section*{Funding}
This research received no external funding.
\section*{Disclosure statement}
The author reports no conflict of interest.
\section*{Use of artificial intelligence}
During manuscript preparation, ChatGPT (OpenAI) was used for English-language
editing and stylistic refinement. The author reviewed all content and takes
full responsibility for the accuracy and scientific integrity of the manuscript.
\section*{Data availability}
The QSAR aquatic toxicity data are publicly available from the UCI Machine
Learning Repository at \url{https://doi.org/10.24432/C5SG7H}.
\section*{Supplementary material}
An anonymous technical supplement contains the extended penalty family,
the proofs of the main results, and the local correlation analysis.

\FloatBarrier

\end{document}


\spacingset{1}
\bigskip
\begin{center}{\LARGE\bfseries Supplementary material: The Lambert Penalty: Logarithmic Shrinkage for Sparse Regression}\end{center}
\noindent This supplement gives the analytic family, technical proofs, and local correlation analysis. Unprefixed references refer to the main article.
\spacingset{1}
\section{General family and analytic construction}
\label{sec:family}

Let $y_n\in\R^n$ be a response vector, let $X_n\in\R^{n\times p}$ be a
design matrix, and let $\beta\in\R^p$ be a coefficient vector.  For a penalty
level $\lambda>0$ and an auxiliary shape parameter $c\geq 1$, the extended estimator is
any global minimizer of the criterion
\begin{equation}
 Q_{n,\lambda,c}(\beta)
 =\frac{1}{2n}\lVert y_n-X_n\beta\rVert_2^2
  +\sum_{j=1}^p p_{\lambda,c}(|\beta_j|).
 \label{eq:criterion}
\end{equation}
Equation~\eqref{eq:criterion} defines the statistical estimator studied in
this section.  Its scalar normalization assumes unit coordinate curvature;
therefore, a coordinate update uses the threshold derived below only after
the corresponding predictor has been standardized to have squared norm $n$.

The main penalty is defined directly in Equation~\exteqref{eq:mainpenalty}.
This appendix records the larger analytic family; its shape is not an
additional tuning parameter of the main method.

The construction begins with a desired threshold rule and then identifies its
induced penalty.  Such inverse constructions are established in the proximal
literature \citep{chartrand2014,woodworthchartrand2016}.  Our proposed design
restriction and the particular rule that saturates it are the objects under
study.  The formula audit in Section~\ref*{sec:audit} is deliberately
bounded and does not establish that no earlier publication contains an
equivalent reparameterization.

\subsection{Derivative-budget construction}
\label{sec:construction}

Fix $\lambda>0$ and $c\geq1$.  On the positive half-line, consider every
absolutely continuous retention rule $T:[\lambda,\infty)\to[0,\infty)$ that
satisfies $T(\lambda)=0$ and $0\leq T(z)\leq z$.  We additionally restrict its
almost-everywhere derivative by
\begin{equation}
 T'(z)\leq \frac{1}{c}+\frac{T(z)}{z},
 \qquad z>\lambda.
 \label{eq:budget}
\end{equation}
Equation~\eqref{eq:budget} is the proposed derivative budget.  The term
$T(z)/z$ makes the restriction invariant to a common rescaling of $z$ and
$\lambda$, while $1/c$ controls how rapidly a coefficient can enter after the
threshold.  It is a design premise, rather than a consequence of risk
optimality.

\begin{proposition}[Maximum retention under the derivative budget]
Within the class just defined, the pointwise largest feasible odd threshold
rule is
\begin{equation}
 T_{\lambda,c}(z)=
 \begin{cases}
 0, & |z|\leq\lambda,\\[2pt]
 \displaystyle \frac{z}{c}\log\!\left(\frac{|z|}{\lambda}\right),
     & \lambda<|z|<e^c\lambda,\\[7pt]
 z, & |z|\geq e^c\lambda.
 \end{cases}
 \label{eq:threshold}
\end{equation}
Equation~\eqref{eq:threshold} defines the Lambert shrinkage rule.  Its
transition begins at zero with right slope $1/c$ and reaches the identity at
the finite cutoff $e^c\lambda$.
\end{proposition}

\begin{proof}
Absolute continuity permits the product rule almost everywhere.  Hence
Equation~\eqref{eq:budget} implies
$(T(z)/z)'=(zT'(z)-T(z))/z^2\leq1/(cz)$.  Integrating from $\lambda$ to $z$
and using $T(\lambda)=0$ gives
$T(z)\leq(z/c)\log(z/\lambda)$.  The constraint $T(z)\leq z$ gives the
second upper bound.  Their minimum is the positive part of
Equation~\eqref{eq:threshold}: the logarithmic branch attains
Equation~\eqref{eq:budget} with equality, and the identity branch remains
feasible because $1\leq1+1/c$.  Odd extension completes the rule.
\end{proof}

Let $W_0:[0,\infty)\to[0,\infty)$ denote the principal real Lambert function,
defined by $W_0(x)e^{W_0(x)}=x$.  For a coefficient magnitude $t\geq0$, the
bounded auxiliary variable
\begin{equation}
 w_{\lambda,c}(t)
 =\min\!\left\{W_0\!\left(\frac{ct}{\lambda}\right),c\right\}
 \label{eq:wdef}
\end{equation}
in Equation~\eqref{eq:wdef} inverts the transition in
Equation~\eqref{eq:threshold}: before the cutoff,
$t=\lambda w e^w/c$ and the corresponding unthresholded score is
$z=\lambda e^w$.

Set $p_{\lambda,c}(0)=0$.  Integrating the first-order condition
$z=t+p'_{\lambda,c}(t)$ along that inverse curve produces
\begin{equation}
 p_{\lambda,c}(t)
 =\frac{\lambda^2}{4c^2}
 \left[
 e^{2w}\{c(1+2w)-2w^2\}-c
 \right],
 \qquad w=w_{\lambda,c}(t).
 \label{eq:penalty}
\end{equation}
Equation~\eqref{eq:penalty} defines the nonnegative penalty used in
Equation~\eqref{eq:criterion}.  The normalization $p_{\lambda,c}(0)=0$
fixes the otherwise irrelevant additive constant.

On the open transition $0<t<e^c\lambda$, differentiation of
Equation~\eqref{eq:penalty}, together with
$dt=\lambda e^w(1+w)\,dw/c$, yields
\begin{equation}
 p'_{\lambda,c}(t)=\lambda e^w\left(1-\frac{w}{c}\right),
 \qquad
 p''_{\lambda,c}(t)=\frac{c}{1+w}-1,
 \qquad
 p'''_{\lambda,c}(t)
 =-\frac{c^2}{\lambda e^w(1+w)^3}<0.
 \label{eq:derivatives}
\end{equation}
Equation~\eqref{eq:derivatives} describes the slope and curvature that govern
sparsity, shrinkage bias, and local optimization.  Here and below, $w$ means
$W_0(ct/\lambda)$ whenever $t$ is strictly inside the transition.  At the
origin, the one-sided limits are $p'_{\lambda,c}(0+)=\lambda$ and
$p''_{\lambda,c}(0+)=c-1$.

At and beyond the identity cutoff, Equation~\eqref{eq:penalty} becomes
\begin{equation}
 p_{\lambda,c}(t)=C_c\lambda^2,
 \qquad
 C_c=\frac{e^{2c}-1}{4c},
 \qquad
 p'_{\lambda,c}(t)=0,
 \qquad t\geq e^c\lambda.
 \label{eq:tail}
\end{equation}
Equation~\eqref{eq:tail} identifies the finite flat tail that removes
penalization bias for sufficiently large coefficients.  The first derivative
is continuous at the cutoff.  The second derivative is undefined there: its
left limit is $-1/(c+1)$ and its right limit is zero.

The smallest transition curvature in Equation~\eqref{eq:derivatives} is the
left limit at the cutoff.  Including the upward derivative jump at zero gives
the global weak-convexity statement
\begin{equation}
 p''_{\lambda,c}(t)\geq-\frac{1}{c+1}
 \quad\text{wherever it exists},
 \qquad
 b\longmapsto p_{\lambda,c}(|b|)+\frac{b^2}{2(c+1)}
 \quad\text{is convex}.
 \label{eq:weakconvex}
\end{equation}
Equation~\eqref{eq:weakconvex} supplies a uniform curvature bound for the
scalar problem.  In particular, adding the unit-curvature squared loss makes
the scalar objective $c/(c+1)$-strongly convex.

\begin{theorem}[Unique scalar update]
\label{thm:familyscalar}
For every $z\in\R$, $\lambda>0$, and $c\geq1$, the unique minimizer of
$b\mapsto (z-b)^2/2+p_{\lambda,c}(|b|)$ is
$T_{\lambda,c}(z)$ from Equation~\eqref{eq:threshold}.
\end{theorem}

\begin{proof}
Strong convexity follows from Equation~\eqref{eq:weakconvex}, so it is enough
to verify the first-order condition.  At zero, the one-sided penalty slope
gives the subgradient condition $|z|\leq\lambda$.  In the positive transition,
Equations~\eqref{eq:wdef} and \eqref{eq:derivatives} give
$t+p'_{\lambda,c}(t)=\lambda e^w=z$; odd symmetry handles negative scores.
In the tail, Equation~\eqref{eq:tail} gives $p'_{\lambda,c}(t)=0$, so the
minimizer is $b=z$.  These three cases reproduce
Equation~\eqref{eq:threshold}.
\end{proof}

For $c>1$, the curvature in Equation~\eqref{eq:derivatives} is positive near
the origin precisely on the interval
\begin{equation}
 0<t<a_c\lambda,
 \qquad
 a_c=\frac{(c-1)e^{c-1}}{c}.
 \label{eq:positivecurvature}
\end{equation}
Equation~\eqref{eq:positivecurvature} locates the region in which the penalty
adds local curvature to the loss.  Increasing $c$ lowers the threshold's entry
slope to $1/c$, but also moves the identity cutoff to $e^c\lambda$; local
stabilization is therefore purchased with a longer shrinkage region.  At
$c=1$, Equation~\eqref{eq:positivecurvature} is empty, and the formula reduces
to the original $c=1$ Lambert rule.

\section{Uniform oracle equivalence}
\label{sec:oracle}

We now show that the boundary shape $c=1$ does not have to be excluded from a
fixed-dimensional oracle result.  The proof uses a uniform positive lower
bound on the penalty slope in a shrinking neighborhood of zero, rather than
the positive-curvature interval in Equation~\eqref{eq:positivecurvature}.

Assume the fixed-dimensional linear model
$y_n=X_n\beta^0+\varepsilon_n$, where $p$ and
$\beta^0\in\R^p$ are fixed, $X_n$ is deterministic,
$G_n=X_n^\mathsf{T}X_n/n\to\Sigma\succ0$, and
$\varepsilon_n\sim N(0,\sigma^2I_n)$ for a fixed
$0<\sigma^2<\infty$.  Let $S=\supp(\beta^0)$ and $N=S^c$; each nonzero
component of $\beta^0$ has fixed positive magnitude.  Fix a finite constant
$C\geq1$.  For deterministic positive sequences
$\underline\lambda_n\leq\overline\lambda_n$, define the tuning set
\begin{equation}
 \Theta_n=[\underline\lambda_n,\overline\lambda_n]\times[1,C],
 \qquad
 \overline\lambda_n\to0,
 \qquad
 \sqrt n\,\underline\lambda_n\to\infty.
 \label{eq:band}
\end{equation}
Equation~\eqref{eq:band} separates the two roles of the penalty level: the
upper endpoint makes shrinkage vanish on fixed signals, and the lower endpoint
dominates the $n^{-1/2}$ noise score.  No ratio condition between the two
endpoints is required.

For nonempty $S$, define the oracle least-squares estimator by fitting only the
true support; for empty $S$, define it as zero.  In symbols,
\begin{equation}
 \widehat\beta^{\mathrm{or}}_S
 =(X_{n,S}^{\mathsf T}X_{n,S})^{-1}X_{n,S}^{\mathsf T}y_n,
 \qquad
 \widehat\beta^{\mathrm{or}}_N=0.
 \label{eq:oracleestimator}
\end{equation}
Equation~\eqref{eq:oracleestimator} is the benchmark that knows the true
support before observing the response.

\begin{theorem}[Uniform global-oracle equivalence]
\label{thm:familyoracle}
Under the preceding assumptions and Equation~\eqref{eq:band}, with probability
tending to one, every global minimizer of
$Q_{n,\lambda,c}$ equals $\widehat\beta^{\mathrm{or}}$ simultaneously for
all $(\lambda,c)\in\Theta_n$.  Consequently, if a possibly data-dependent
tuning pair $(\widehat\lambda_n,\widehat c_n)$ belongs to $\Theta_n$ with
probability tending to one, then every global minimizer selected at that pair
recovers $S$ consistently and, when $S$ is nonempty,
\begin{equation}
 \sqrt n\left(\widehat\beta_S-\beta^0_S\right)
 \ \Rightarrow\ 
 N\!\left(0,\sigma^2\Sigma_{SS}^{-1}\right).
 \label{eq:oraclelimit}
\end{equation}
Equation~\eqref{eq:oraclelimit} states that the active estimates have the same
first-order distribution as least squares computed with the true support.
\end{theorem}

\begin{proof}
For all sufficiently large $n$, $G_n\succeq\delta I_p$ for a constant
$\delta>0$, which also makes Equation~\eqref{eq:criterion} coercive.  Thus
global minimizers exist.  Put
$r_n=X_n^{\mathsf T}\varepsilon_n/n$ and
$\overline C=\max_{1\leq c\leq C}C_c<\infty$, where $C_c$ is defined in
Equation~\eqref{eq:tail}.  Comparing any global minimizer
$\widehat\beta^{\lambda,c}$ with $\beta^0$ gives the simultaneous bound
\begin{equation}
 \frac{\delta}{2}\lVert h\rVert_2^2
 \leq \lVert r_n\rVert_2\lVert h\rVert_2
      +p\overline C\,\overline\lambda_n^2,
 \qquad
 h=\widehat\beta^{\lambda,c}-\beta^0,
 \qquad (\lambda,c)\in\Theta_n.
 \label{eq:localization}
\end{equation}
Equation~\eqref{eq:localization} localizes every global minimizer uniformly,
because the penalty is nonnegative and is bounded above by
$\overline C\overline\lambda_n^2$ per coordinate.  Since
$\lVert r_n\rVert_2=O_p(n^{-1/2})$, solving the quadratic inequality in
Equation~\eqref{eq:localization} shows that the supremum of
$\lVert h\rVert_2$ over all tuning pairs and all global minimizers is
$O_p(n^{-1/2}+\overline\lambda_n)=o_p(1)$.

The oracle estimator in Equation~\eqref{eq:oracleestimator} is also
consistent.  Fixed nonzero signals, uniform localization, and
$e^C\overline\lambda_n\to0$ imply that, on an event whose probability tends
to one, every active coordinate of every global minimizer and of the oracle
estimator lies strictly in the flat region in Equation~\eqref{eq:tail}.  On
this event the active penalty derivatives vanish.  If $N$ is empty, active
stationarity immediately gives the oracle estimator, so suppose $N$ is
nonempty.

Let $M_S=I_n-X_{n,S}(X_{n,S}^{\mathsf T}X_{n,S})^{-1}X_{n,S}^{\mathsf T}$,
with $M_S=I_n$ when $S$ is empty. Let
$v=\widehat\beta_N^{\lambda,c}$ and define
$H_n=X_{n,N}^{\mathsf T}M_SX_{n,N}/n$ and
$R_n=X_{n,N}^{\mathsf T}M_S\varepsilon_n/n$.  Active stationarity first
profiles out $\widehat\beta_S^{\lambda,c}$.  Because the active penalties of
the global and oracle estimators are then equal constants, their objective
difference is
\begin{equation}
 Q_{n,\lambda,c}(\widehat\beta^{\lambda,c})
 -Q_{n,\lambda,c}(\widehat\beta^{\mathrm{or}})
 =\frac12v^{\mathsf T}H_nv-R_n^{\mathsf T}v
  +\sum_{j\in N}p_{\lambda,c}(|v_j|).
 \label{eq:profile}
\end{equation}
Equation~\eqref{eq:profile} isolates the inactive coordinates.  The Schur
complement satisfies $H_n\succeq\delta I$, and fixed dimension gives
$\lVert R_n\rVert_2=O_p(n^{-1/2})$.

Choose any fixed $w_0\in(0,1)$ and define
$a_*=w_0e^{w_0}/C$ and $\kappa=1-w_0$.  If
$0\leq t\leq a_*\lambda$ and $c\in[1,C]$, then
$W_0(ct/\lambda)\leq w_0$.  Equation~\eqref{eq:derivatives} therefore yields
the uniform local bound
\begin{equation}
 p'_{\lambda,c}(t)\geq\kappa\lambda,
 \qquad
 p_{\lambda,c}(t)\geq\kappa\lambda t,
 \qquad
 0\leq t\leq a_*\lambda,\quad c\in[1,C].
 \label{eq:localslope}
\end{equation}
Equation~\eqref{eq:localslope} is the step that includes $c=1$: it needs a
positive slope near zero, but not positive penalty curvature. The derivative
at zero in Equation~\eqref{eq:localslope} denotes the right derivative.

With probability tending to one,
$\lVert R_n\rVert_\infty<\kappa\underline\lambda_n$ and
$2\lVert R_n\rVert_2/\delta<a_*\underline\lambda_n$, by the lower-band
condition in Equation~\eqref{eq:band}.  If
$\lVert v\rVert_2>2\lVert R_n\rVert_2/\delta$, the quadratic and linear terms
in Equation~\eqref{eq:profile} have a strictly positive sum.  Otherwise every
$|v_j|$ lies in the range of Equation~\eqref{eq:localslope}, and
Equation~\eqref{eq:profile} is at least
$(\kappa\underline\lambda_n-\lVert R_n\rVert_\infty)\lVert v\rVert_1$,
which is strictly positive for $v\neq0$.  A global minimizer cannot have a
larger objective than the oracle estimator; hence $v=0$.  Active stationarity
then gives Equation~\eqref{eq:oracleestimator}.  The same events work for all
$(\lambda,c)\in\Theta_n$, proving simultaneous equivalence.  Finally,
Gaussian least-squares theory and $G_{n,SS}\to\Sigma_{SS}$ give
Equation~\eqref{eq:oraclelimit}.
\end{proof}

\begin{remark}[What the theorem does not certify]
The theorem concerns fixed $p$, fixed nonzero signals, deterministic designs,
global minimizers, and tuning pairs inside Equation~\eqref{eq:band}.  It does
not establish that ordinary cross-validation selects a penalty level in that
band.  It also does not cover $p$ growing with $n$, correlations converging to
one, local stationary points returned by an algorithm, or uniform numerical
convergence.  These distinctions are material because nonconvex objectives
may admit several stationary solutions \citep{brehenyhuang2011}.
\end{remark}

\section{What local correlation analysis can establish}
\label{sec:correlation}

The positive-curvature regime below belongs to the extended family with
$c>1$, not to the main Lambert penalty. For the main method,
Equation~\exteqref{eq:mainderivatives} gives nonpositive curvature, so the
positive-curvature grouping improvement is unavailable. This section states
a conditional diagnostic, not an advantage claimed for the empirical method.

Consider two standardized predictors with Gram matrix
$G=\left(\begin{smallmatrix}1&\rho\\\rho&1\end{smallmatrix}\right)$,
where $0\leq\rho<1$, and score vector $g\in\R^2$.  After discarding a constant,
the two-coordinate objective is
\begin{equation}
 L(\beta_1,\beta_2)
 =\frac12\begin{pmatrix}\beta_1&\beta_2\end{pmatrix}
 G\begin{pmatrix}\beta_1\\\beta_2\end{pmatrix}
 -g^{\mathsf T}\begin{pmatrix}\beta_1\\\beta_2\end{pmatrix}
 +\sum_{j=1}^2p_{\lambda,c}(|\beta_j|).
 \label{eq:twoobjective}
\end{equation}
Equation~\eqref{eq:twoobjective} is a local diagnostic model, rather than a
claim about the global fit in Equation~\eqref{eq:criterion}.

At a stationary point with both coefficients nonzero and away from the joins,
write $h_j=p''_{\lambda,c}(|\beta_j|)$.  The local Hessian and the Jacobian of
one full cycle that updates coordinate 1 and then coordinate 2 are
\begin{equation}
 \begin{gathered}
 H_{\mathrm{loc}}
 =\begin{pmatrix}1+h_1&\rho\\\rho&1+h_2\end{pmatrix},
 \qquad
 J_{\mathrm{cyc}}
 =\begin{pmatrix}
 0&-\rho/(1+h_1)\\
 0&\rho^2/\{(1+h_1)(1+h_2)\}
 \end{pmatrix},
 \\[3pt]
 q=\frac{\rho^2}{(1+h_1)(1+h_2)}.
 \end{gathered}
 \label{eq:cycle}
\end{equation}
Equation~\eqref{eq:cycle} follows by implicit differentiation of each unique
scalar update. Its eigenvalues are $0$ and $q$. If the local Hessian is
positive definite, then $q<1$ and the differentiable local update map is
locally attracting. A strict local minimum alone does not imply a positive
definite Hessian; a degenerate point with $q=1$ is not covered by this
linear contraction argument. If $q>1$, the
stationary point is unstable in the coordinate map; $q$ is then not a
convergence rate for a valid fitted solution.

If $h_1,h_2\geq m>0$, Equation~\eqref{eq:cycle} gives
$q\leq\rho^2/(1+m)^2$ and
$\lambda_{\min}(H_{\mathrm{loc}})\geq1-\rho+m$.  For two positive
coefficients, assume additionally that
$p''_{\lambda,c}(t)\geq m$ throughout the interval between their magnitudes.
Subtracting their stationarity equations and applying the mean-value theorem
then gives
\begin{equation}
 |\beta_1-\beta_2|
 \leq\frac{|g_1-g_2|}{1-\rho+m}.
 \label{eq:grouping}
\end{equation}
Equation~\eqref{eq:grouping} is a local grouping bound: extra positive
curvature reduces sensitivity of the coefficient contrast to a score
contrast.  It requires a common positive sign and a curvature lower bound on
the entire intervening interval.

The benefit is regime dependent.  When both coefficients enter the flat tail
in Equation~\eqref{eq:tail}, $h_1=h_2=0$, so
Equation~\eqref{eq:cycle} returns to $q=\rho^2$ and
Equation~\eqref{eq:grouping} loses its curvature term.  For identical columns,
the data identify only a sum of coefficients; a penalty can select an
allocation but cannot restore the missing information.  Thus the analysis
supports local stabilization in the positive-curvature region, not a general
solution to collinearity.  Correlated-variable penalties with grouping and
oracle motivations already include the elastic net, adaptive elastic net, and
Mnet \citep{zouhastie2005,zouzhang2009,huangetal2016}.

\section{Proofs of the main results}
\label{sec:mainproofs}
\subsection{Proof of Proposition~\ref*{prop:mainprox}}
\label{sec:proof1}
\begin{proof}
At $c=1$, Equation~\eqref{eq:weakconvex} establishes convexity of
$b\mapsto p_\lambda(|b|)+b^2/4$ and $1/2$-strong convexity of the scalar
objective. The constant is sharp: Equation~\exteqref{eq:mainderivatives}
gives $p''_\lambda(t)\to-1/2$ as $t\uparrow e\lambda$. For any proposed
weak-convexity constant $a<1/2$, the adjusted curvature
$p''_\lambda(t)+a$ is therefore negative sufficiently close to that cutoff,
so $b\mapsto p_\lambda(|b|)+ab^2/2$ cannot be convex.
Theorem~\ref{thm:familyscalar}, specialized to $c=1$, gives the unique
minimizer $T_\lambda(z)$ in Equation~\exteqref{eq:mainthreshold}.
\end{proof}
\subsection{Proof of Theorem~\ref*{thm:globalunique}}
\label{sec:proof2}
\begin{proof}
Let $g=X^{\mathsf T}y/n$ and $C_y=\|y\|_2^2/(2n)$. By adding and
subtracting $\|\beta\|_2^2/4$, the criterion becomes
\begin{equation*}
 Q_\lambda(\beta)=C_y-g^{\mathsf T}\beta
 +\frac12\beta^{\mathsf T}(G-I_p/2)\beta
 +\sum_{j=1}^p\left\{p_\lambda(|\beta_j|)+\frac{\beta_j^2}{4}\right\}.
\end{equation*}
This decomposition separates a quadratic with the stated
curvature from convex coordinate penalties, by Proposition~\ref*{prop:mainprox}.
It proves both convexity assertions. For $\mu>0$ its quadratic lower bound
is coercive, ensuring existence; strong convexity gives uniqueness and the
stationarity conclusion. No uniqueness claim is made at $\mu=0$.
\end{proof}
\subsection{Proof of Theorem~\ref*{thm:cd}}
\label{sec:proof3}
\begin{proof}
Holding all but one coordinate fixed gives the scalar objective of
Proposition~\ref*{prop:mainprox}, up to a constant. Its strong convexity gives
the first inequality in Equation~\exteqref{eq:maindescent}. Since $Q_\lambda$
is nonnegative, summing the decreases proves square summability and
$\|b^{k+1}-b^k\|_2\to0$.

Suppose a subsequence converges to $b^*$. The next at most $p$ iterates have
the same limit because their increments vanish. For every coordinate, choose
its update in this finite cycle. The partial-residual score depends
continuously on the other coordinates, and $T_\lambda$ is continuous,
including its two joins. Passing to the limit in that update proves that
$b^*$ minimizes each coordinate objective. Its one-dimensional optimality
conditions are precisely the stationarity conditions stated above.

For bounded iterates, the accumulation set is nonempty and compact, and the
distance of the iterates to that set tends to zero. If stationary points in
the initial sublevel set are finite, enclose the possible accumulation points
in disjoint small balls separated by a positive distance. Eventually all
iterates lie in their union, and vanishing increments prevent moves between
balls. Hence there is only one accumulation point, and the sequence converges.
With full column rank, $G$ is positive definite and the nonnegative penalty
makes the initial sublevel set compact. Finally, under
Theorem~\ref*{thm:globalunique} this bounded sequence has only one possible
stationary accumulation point, the unique global minimizer.
\end{proof}
\subsection{Proof of Theorem~\ref*{thm:mainoracle}}
\label{sec:proof4}
\begin{proof}
Apply Theorem~\ref{thm:familyoracle} with $C=1$, so the tuning set in
Equation~\eqref{eq:band} fixes $c=1$ and uses the penalty band in
Equation~\exteqref{eq:mainband}. Its assumptions then coincide with those of
Theorem~\ref*{thm:mainoracle}. The simultaneous oracle equality, support
consistency, and active-coordinate limit in
Equation~\exteqref{eq:mainoraclelimit} follow directly, including the stated
allowance for data-dependent tuning within the band.
\end{proof}

\FloatBarrier